\documentclass{article}
\usepackage{tempfras}

\def\mdmatm{\Delta m^2_{32}}
\def\dmatm{$\mdmatm$}
\def\mdmsol{\Delta m^2_{21}}

\def\meV{e\mbox{V}}

\begin{document}
\title{ 
The Case for a Super Neutrino Beam
}
\author{
Milind V. Diwan \\
{\em Brookhaven National Laboratory} 
}
\maketitle
\baselineskip=11.6pt
\begin{abstract}
In  this paper I will discuss how an
intense beam of high energy neutrinos 
produced with  conventional technology 
 could be used to further our understanding of neutrino masses 
and mixings. 
 I will describe the possibility of building such a beam at 
existing U.S. laboratories. 
Such a project couples naturally to 
a large ($>$ 100 kT) multipurpose detector in a new deep underground 
laboratory.  I will discuss the requirements for such a detector.
Since the number of sites for both an accelerator laboratory and 
a deep laboratory are limited, I will discuss  how  
the choice of baseline affects the physics sensitivities,  
the practical issues of beam construction, and event rates. 

\end{abstract}
\baselineskip=14pt
\section{Introduction}

\begin{figure}
\vspace{5cm}
\includegraphics{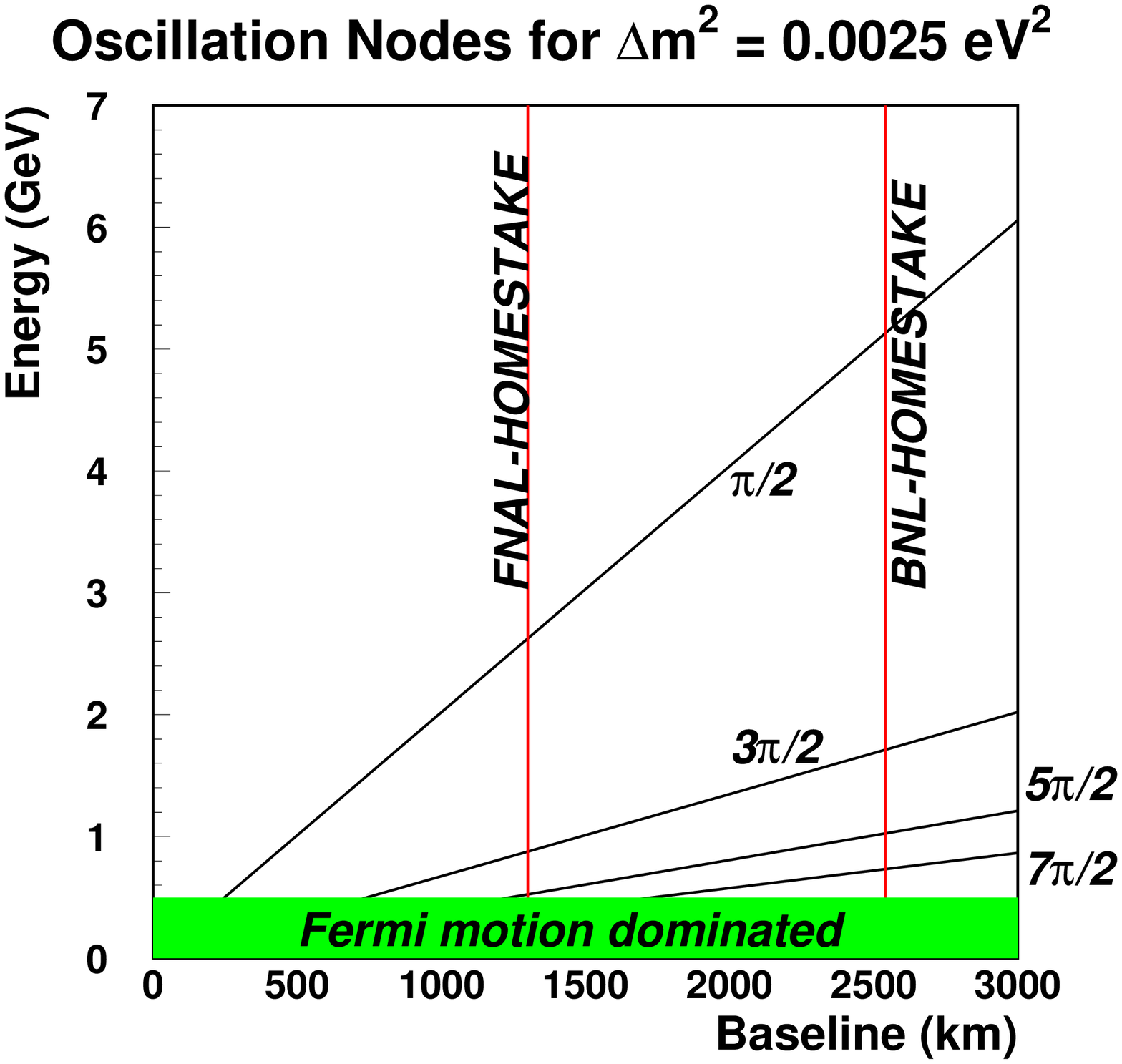}
\includegraphics{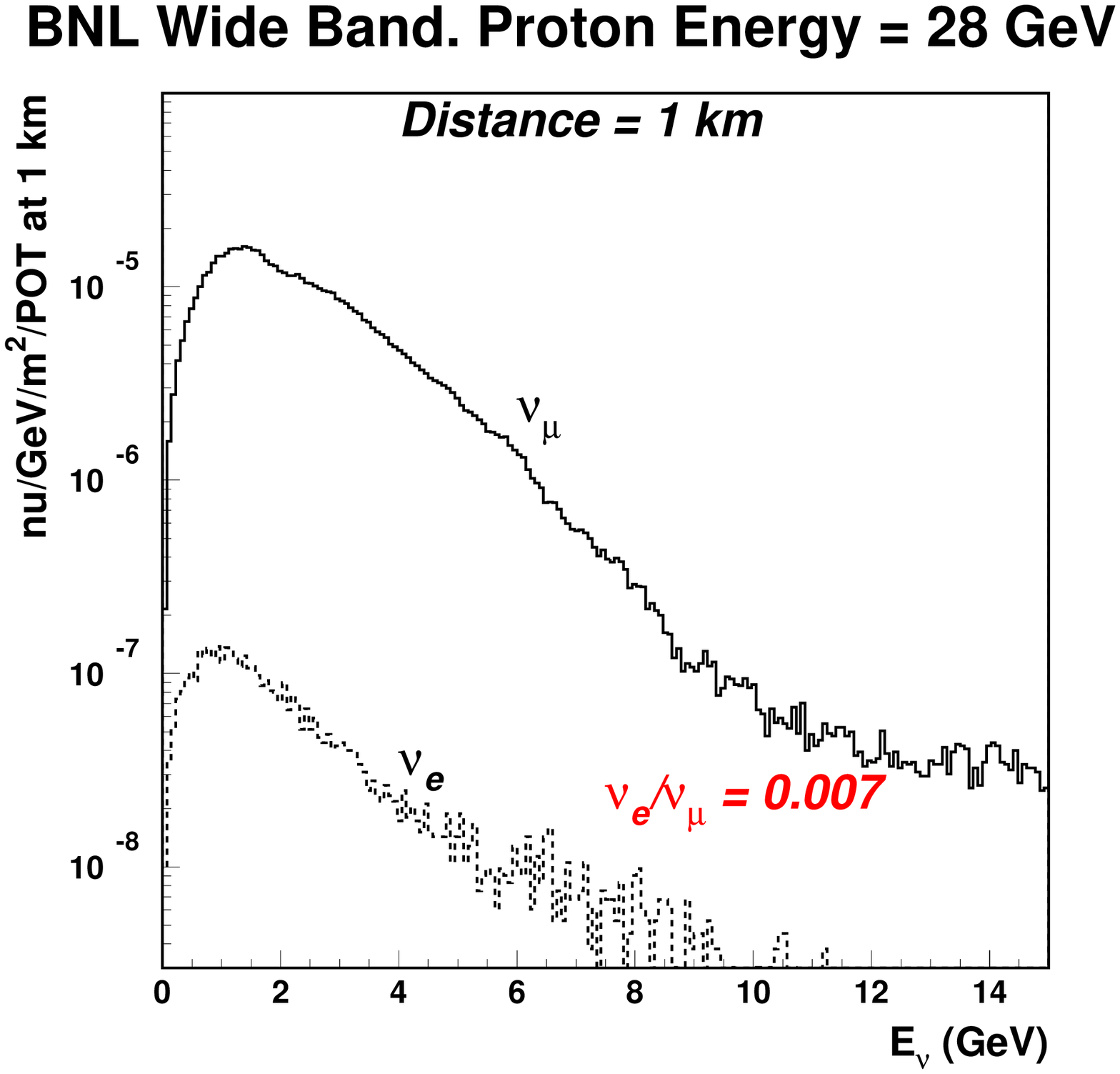}
 \caption{\it
     Nodes of oscillations for $\Delta m^2_{32}=0.0025 eV^2$ in 
neutrino energy versus baseline (left). 
 Possible baselines from Brookhaven National 
Laboratory (BNL) and Fermi National Laboratory (FNAL) to the Homestake 
underground site are indicated.  They correspond to distances of $\sim 2540$ km and 
$\sim 1290$ km, respectively.   
Right hand side shows the wide band neutrino spectrum from 28 GeV protons
at a distance of 1 km from the target. 
The anti-neutrino spectrum looks similar, but has 
contamination from  neutrinos.  
    \label{nodes} }
\end{figure}

In \cite{previous} we argued  that an intense
 broadband muon neutrino beam and a large detector located 
more than 2000 km away from the source could be used to perform 
precision measurements of neutrino properties such as the 
mass differences,   the mass hierarchy, the mixing parameters, and
 CP violation in the neutrino sector.  Using the currently 
deduced neutrino mass differences and mixing parameters
\cite{ref1} and  the same formalism as \cite{previous}
 we formulated several simple rules  for such an experiment: 

For precise measurements of \dmatm and $\sin^2 2 \theta_{23}$, it is
desirable to observe a pattern of multiple nodes in the energy
spectrum of muon neutrinos.  Since the cross section, Fermi motion,
and nuclear effects limit the resolution of muon neutrino interactions
below $\sim$ 1 GeV, we need to utilize a wide band muon neutrino beam
with energy range of 1-6 GeV and a distance of $\sim$2000km to
observe 3 or more oscillation nodes. See Fig. \ref{nodes}.

The appearance spectrum of electron neutrinos from the conversion
$\nu_\mu \to \nu_e$ contains information about $\sin^2 2 \theta_{13}$,
$\delta_{CP}$, $\Delta m_{21}^2$ and the ordering of neutrino masses
through the matter effect (i.e. ($m_1<m_2<m_3$) versus ($m_3<m_1<m_2$)).
We showed that the effects of the various parameters can be separated
using the broad-band 1-6 GeV beam and the $\sim$2000km distance.  The
matter effect causes the conversion probability to rise with energy
and is mostly confined to energies $>$ 3 GeV whereas the effects of
$\delta_{CP}$ fall as $1/E$. We showed that this energy dependence can
be used to measure the value of $\delta_{CP}$ and $\sin^2 2
\theta_{13}$ without taking data with anti-neutrinos.

The additional contribution to the appearance event rate due to 
3-generation CP violation in the neutrino sector 
is approximately proportional to: 
$\sin \delta_{CP}\sin 2 \theta_{13} \times (\Delta m^2_{21} L/4 E_\nu)$.
This contribution increases linearly with distance while the total
flux falls as $1/L^2$ for a detector of a given size. The statistical
sensitivity for the additional CP contribution, however, remains
approximately independent of distance. It is therefore advantageous to
perform the experiment with a very long ($>$ 2000 km) baseline because
then we can relax the requirements on systematic errors on the flux,
the cross sections, the other oscillation parameters, and the calculation
of the matter effect.

Because of the electron neutrino contamination background in a
conventional accelerator neutrino beam the sensitivity to
$\delta_{CP}$ will be limited to the parameter region $\sin^2 2
\theta_{13} > 0.01$. The main CP-conserving contribution to the
$\nu_\mu \to \nu_e$ signal is proportional to $\sin^2 2 \theta_{13}$
in this region.  The CP-violating term, on the other hand, is linear
in $\sin 2 \theta_{13}$.  Therefore the fractional contribution due to
the CP-violating term increases for small $\sin 2 \theta_{13}$,
although the total appearance signal decreases. The statistical
sensitivity to the CP-violating term remains approximately independent
of the value of $\sin^2 2 \theta_{13}$ as long as backgrounds do not
dominate the observed spectrum\cite{ref3}.  When $\sin^2 2
\theta_{13}$ is very small ($<0.002$) this rule no longer holds
because the signal is no longer dominated by the $\sin^2 2
\theta_{13}$ term in the 3-generation formalism
\cite{ref4}.   

Current generation of accelerator experiments such as K2K \cite{k2k},
MINOS \cite{minos}, or CNGS \cite{CNGS} focus on obtaining a
definitive signature of muon neutrino oscillations at the first node
($\Delta m^2_{32} L/4E \sim \pi/2$) for the atmospheric mass
scale. Other recent proposed projects (JPARC-to-SK, NUMI-offaxis) 
\cite{offaxis,numioff} also focus mainly on the first node, but propose
to use an off-axis narrow band beam to lower the background in the
search of $\nu_\mu \to \nu_e$ caused by a non-zero $\theta_{13}$. The
narrow band beam and limited statistics, however, do not allow
measurement of the parameters in a definitive way.  Proposed reactor
disappearance searches, also at the first node for the atmospheric
mass scale, are only sensitive to $\sin^2 2 \theta_{13}$ \cite{ref4}.

Thus, current and near term accelerator based experiments are focussed
on the atmospheric mass scale.  Experiments using astrophysical
sources such as solar neutrinos or atmospheric neutrinos are sensitive
to either the solar or the atmospheric mass scale. The parameters are
now known well enough ($\Delta m^2_{32}\sim 0.0025 eV^2$ and $\Delta
m^2_{21} \sim 8\times 10^{-5} eV^2$ ) \cite{sk, kamland, sno} that it
is possible to design a qualitatively different experiment that will
have good sensitivity to both mass scales. The CP
contribution is dependent on both atmospheric and solar $\Delta m^2$;
it is also likely that such an experiment is necessary to uncover any
new physics in neutrino mixing or interactions with matter.  A next
generation accelerator experiment with well understood, pure beams,
sufficiently long baseline, and low energy wide band beam (1-5 GeV)
could fill this role.

In this paper we will discuss different options for the baseline. In
\cite{previous} we demonstrated that for 3-generation mixing the CP
parameters could be measured using neutrino data alone. Any additional
information from anti-neutrino running therefore could 
make the measurements more precise as well as constrain
contributions from new physics, in particular, new interactions in
matter or new sources of CP violation in the neutrino sector.
 We will calculate the significance with which the neutrino mass and mixing
parameters can be measured using both neutrino and anti-neutrino data
and the implications for the determination of the mass hierarchy and
demonstration of CP violation.

\section{Accelerator and Detector Requirements}

Previously we described  the BNL Alternating 
Gradient Synchrotron 
(AGS)  operating at 28 GeV
upgraded to provide  total  proton beam 
power of 1 MW \cite{agsup} and a 500 kTon detector placed at the
proposed national underground laboratory (NUSEL) \cite{NUSEL}
in the Homestake mine in South Dakota.
The main components of the accelerator upgrade at BNL are 
a new 1.2 GeV Superconducting LINAC to provide protons to the 
existing AGS, and new magnet power supplies to increase the ramp rate of the 
 AGS magnetic field from about 0.5 Hz  of today to  2.5 Hz.  
For 1 MW operation the protons from the accelerator will be delivered 
in pulses of $9\times 10^{13}$ protons at 2.5 Hz.
We have determined that 2 MW operation of the AGS is also possible by 
further upgrading the synchrotron to 5 Hz repetition rate and with 
further modifications to the LINAC and the RF systems. 
The neutrino beam will be built with conventional horn focussed 
technology and a 200 m long pion decay tunnel.

High energy multi-MW proton beams are also under consideration at
FNAL. The most ambitious plans \cite{foster} call for a 8 GeV
superconducting LINAC that can provide $1.5\times 10^{14}$ $H^-$ ions
at 10 Hz corresponding to 2 MW of total beam power. Some of these 8
GeV ions could be injected into the main injector (MI) to provide 2 MW
proton beam power at any energy between 40 and 120 GeV; for example,
40 GeV at 2 Hz or 120 GeV at 0.67 Hz. Such a plan allows much
flexibility in the choice of proton energy for neutrino production. As
Figure \ref{nodes} shows for observing multiple oscillation nodes in muon
neutrino oscillations it is necessary to have a wide band beam with
energies from 1 to 5 GeV. Protons above $\sim20$ GeV are needed to provide 
such a flux, clearly possible at either BNL or FNAL. 
For the purposes of the analysis in
this paper we will assume that the spectrum from either the BNL or the
FNAL beam will be the same. This will allow us a proper comparison of
the physics issues regarding the baselines.

If a large detector facility (as a part of NUSEL)
\cite{3m,uno,icarus}
 is located at
Homestake (HS) the beam from BNL (FNAL) will have to traverse 2540km
(1290km) through the earth. At BNL the beam would have to be built at
an incline angle of about $11.3^o$.  Current design for such a beam
calls for the construction of a hill with a height of about 50 m
\cite{agsup}.  Such a hill will have the proton target at the top of
the hill and a 200 m long decay tunnel on the downslope. At FNAL the
inclination will be about $5.7^o$.  There is already experience at
FNAL in building the NUMI beam
\cite{minos}; this experience 
could be extended to build  a new beam to HS. In either case, 
it is adequate to have a short decay
tunnel  (200 m) compared to the NUMI tunnel (750 m) to achieve the needed 
flux.
The option of running with a narrow band beam using the off-axis
technique \cite{e889} could be preserved if the decay tunnel is made
sufficiently wide. For example, a 4 m diameter tunnel could allow one
to move and rotate the target and horn assembly so that a $1^o$ 
off-axis beam could be sent to the far detector.

With 1 MW of beam, a baseline of 2540 km, and a 500kT detector we
calculate that we would obtain $\sim$60000 muon charged
current and $\sim$20000 neutral current events for $5\times 10^7 sec$ of
running in the neutrino mode in the absence of oscillations.  For the
same running conditions in the anti-neutrino mode (with the horn
current reversed) we calculate a total of $\sim$19000 anti-muon charged
current and $\sim$7000 neutral current events; approximately 20\% of the
event rate in the anti-neutrino beam will be due to wrong-sign
neutrino interactions.  For the shorter baseline of 1290 km from FNAL
to HS, the event rates will be higher by a factor of $(2540/1290)^2$.
For both neutrino and anti-neutrino running approximately $\sim$0.7\% of the
charged current rate will be from electron charged current events
which form a background to the $\nu_\mu \to \nu_e$ search.  It will be
desirable to obtain similar numbers of events in the anti-neutrino and
the neutrino beam. Therefore, for the calculations in this paper we
assume 1 MW operation for $5\times 10^7 sec$ in the neutrino mode and
2 MW operation for $5\times 10^7 sec$ in the anti-neutrino mode.

A large detector facility at NUSEL will most likely be used for a
broad range of physics goals. Important considerations for such a
detector are the fiducial mass, energy threshold, energy resolution,
muon/electron discrimination, pattern recognition capability, time
resolution, depth of the location, and the cost. Two classes of
detectors are under consideration: water Cherenkov detector
instrumented with photo-multiplier tubes and a liquid Argon based time
projection chamber.

A water Cherenkov detector built in the same manner as the 
super-Kamiokande experiment (with 20 inch photo-multipliers placed on the
inside detector surface covering approximately 40\% of the total area)
\cite{sknim} can achieve the 500 kT mass. This could be done  by  
simply scaling the super-Kamiokande detector to larger size or by
building several detector modules\cite{3m, uno}.  Such a detector
placed underground at NUSEL could have a low energy threshold ($< 10$
MeV), good energy resolution ($< 10\%$) for single particles, good
muon/electron separation ($<$ 1\%), and time resolution ($< $ few ns).
For the experiment we propose here it is important to obtain good
energy resolution on the neutrino energy. This can be achieved in a
water Cherenkov detector by separating quasi-elastic scattering events
with well identified leptons in the final state from the rest of the
charged current events. The fraction of quasi-elastics in the total
charged current rate with the spectrum used in this paper is about
23\% for the neutrino beam and 39\% for the anti-neutrino
beam. Separation of quasi-elastic events from the charged current
background is being used by the K2K experiment \cite{k2k}. Further
work is needed to make this event reconstruction work at higher
energies. The reconstruction algorithm could be enhanced by 
the addition of ring imaging techniques to the detector\cite{ypsilantis}.

A number of proponents have argued that a liquid Argon time projection chamber (LARTPC) could be 
built with total mass approaching 100 kT \cite{icarus}. A fine grained detector such 
as this has much better resolution for separating tracks. It is possible therefore to use a 
large fraction of the charged current cross section 
(rather than only the quasi-elastic events) 
for determining  the neutrino energy 
spectrum. The LARTPC will also have much better particle identification 
capability.  Therefore, a LARTPC with a 
smaller total fiducial mass of $\sim$100 kT than the 500 kT assumed for the water 
Cherenkov tank is expected to have similar performance for the physics. 
 
For the purposes of this paper we will assume the same detector performance as 
described in \cite{previous}. 
For the physics sensitivity calculated in this 
paper we will assume 1 MW operation for $5\times 10^7 sec$ in the neutrino mode
and 2 MW operation for $5\times 10^7 sec$ in the anti-neutrino mode. In both cases 
we will assume a detector fiducial mass of 500 kT. With the running times, the accelerator 
power level, and the detector mass fixed, we will consider two baselines: 1290 km (for 
FNAL to Homestake) and 2540 km (for BNL to Homestake) assuming that the detector 
is located at Homestake.

Lastly, we note that for this analysis the far detector could be at 
several comparable sites in the western US, notably WIPP or the
Henderson mine in Colorado. While the detailed calculations change,
the qualitative results are easily deduced from this work for other 
locations.

\section{$\nu_\mu$ disappearance}

\begin{figure}
\vspace{5cm}
\includegraphics{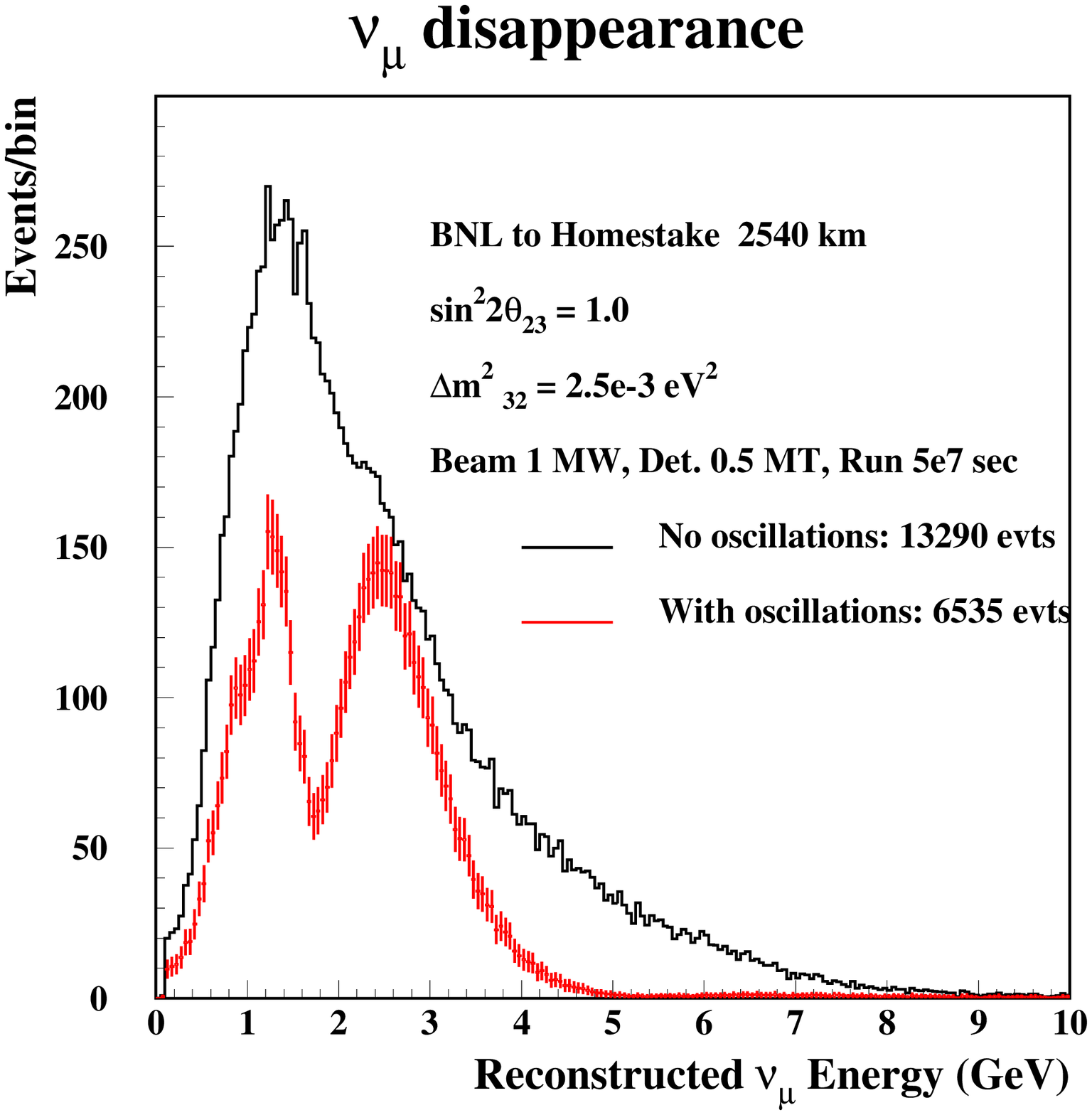}
\includegraphics{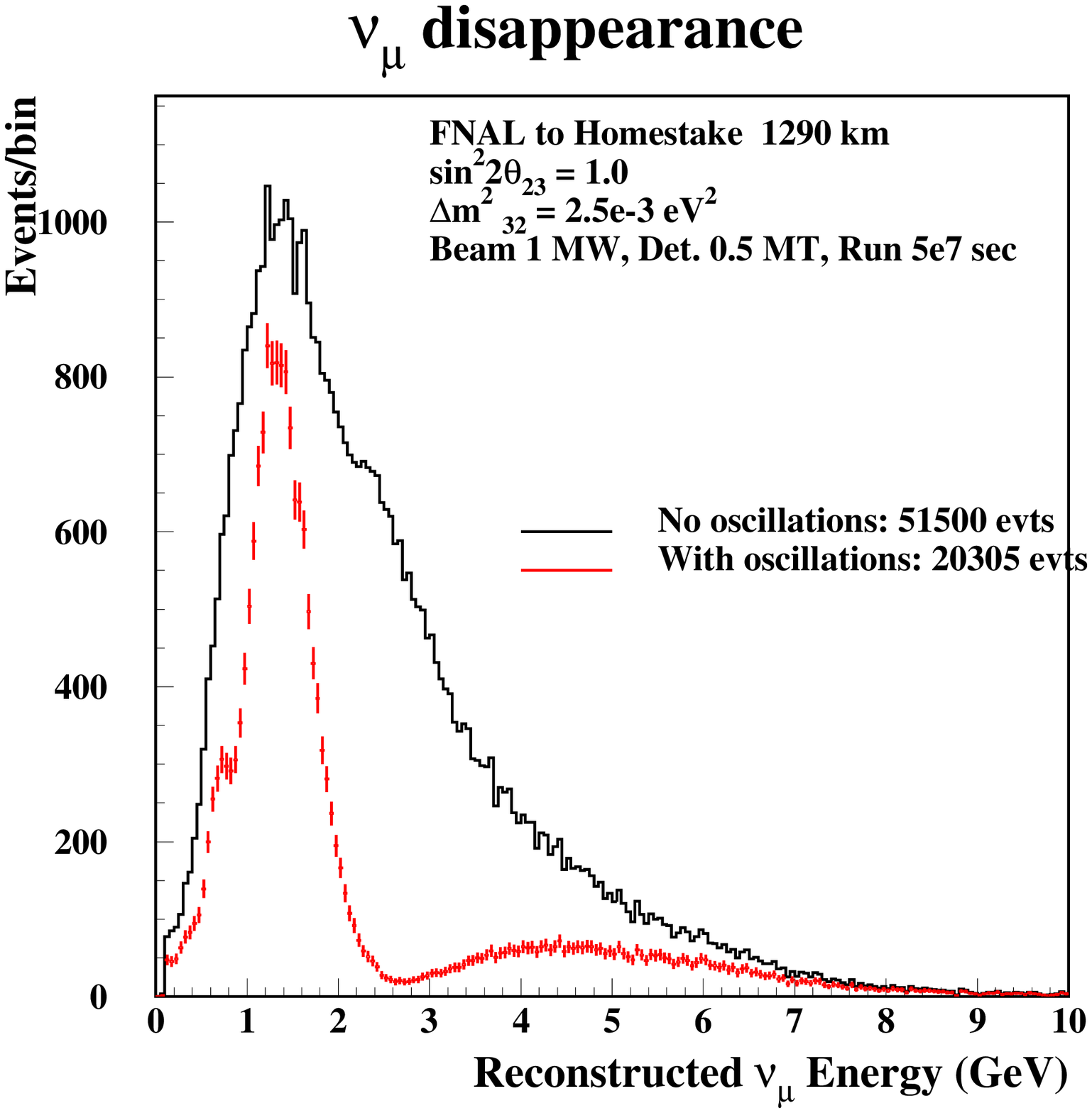}
 \caption{\it
Simulated  spectrum of detected muon neutrinos for 1 MW beam and 500 kT detector 
exposed for $5\times 10^7$ sec. Left side is for baseline of 2540 km, 
right side for baseline of 1290 km. The oscillation parameters assumed 
are shown in the figure.  Only clean single muon events are assumed to be used 
for this measurement (see text).   
    \label{muspec} }
\end{figure}

We propose to use clean single muon events \cite{previous} and
calculate the neutrino energy from the energy and angle of these muons
assuming they are all from quasi-elastic interactions.  The expected
spectrum is shown in Figure \ref{muspec}; the simulation includes
effects of Fermi motion, detector resolution, and backgrounds from
non-quasielastic events.

\begin{figure}
\vspace{4.8cm}
\includegraphics{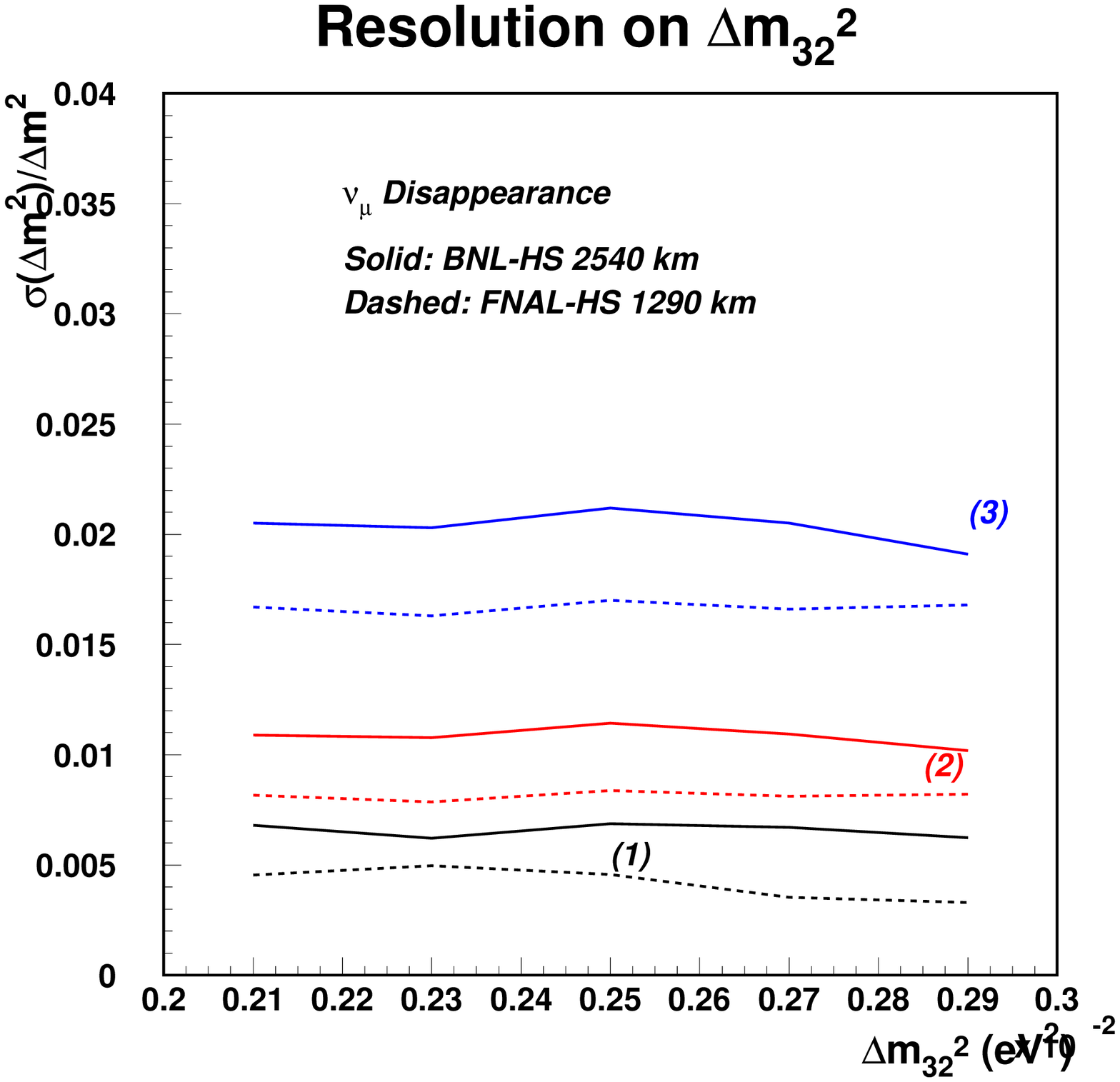}
\includegraphics{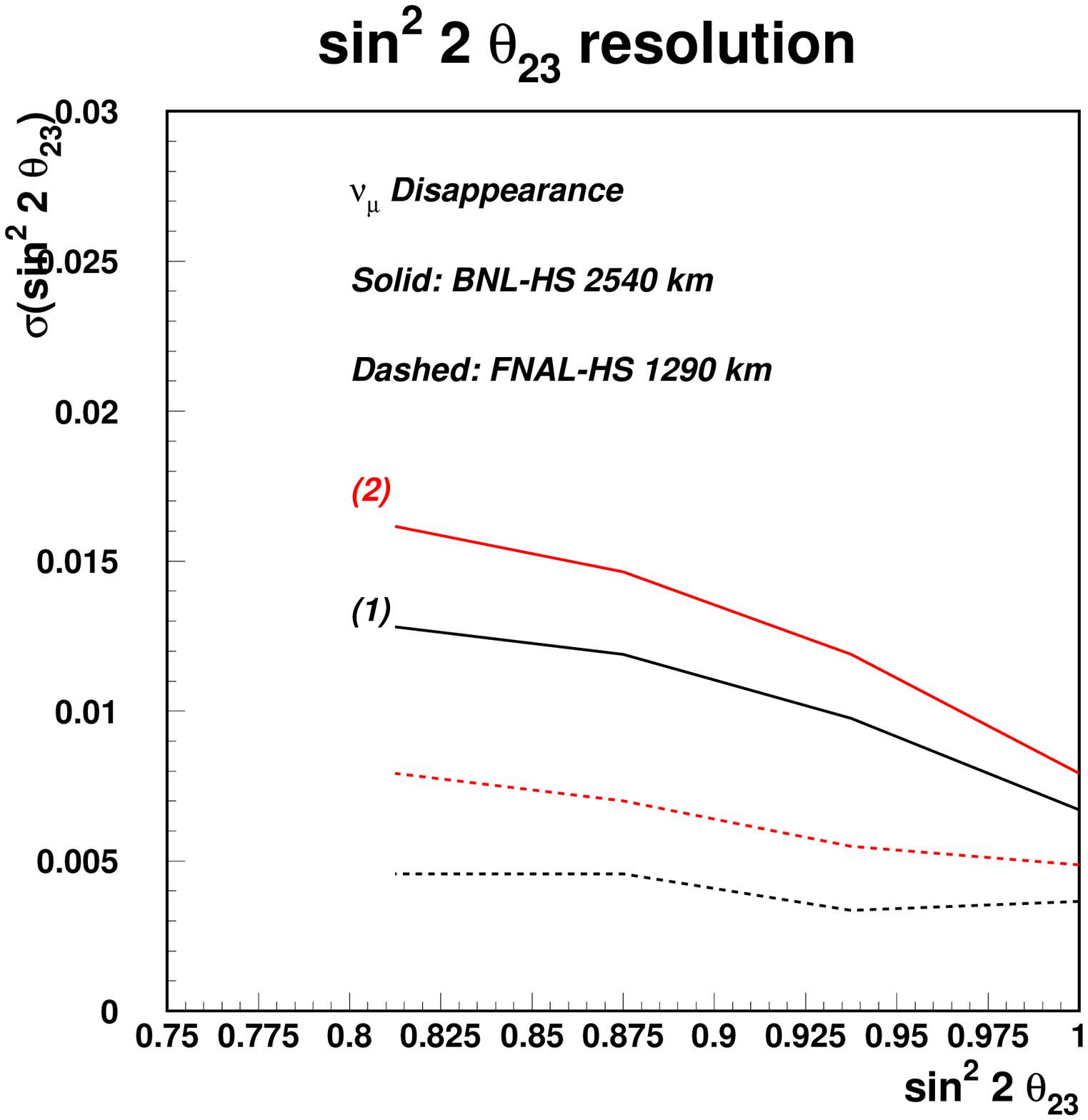}
 \caption{\it 1 sigma resolutions on $\Delta m^2_{32}$ (left) and
 $\sin^2 2 \theta_{23}$ (right) expected after analysis of the
 oscillation spectra from Figure \ref{muspec}. The solid curves are
 for BNL-HS 2540 km baseline, and the dashed are for FNAL-HS 1290 km
 baseline.  The curves labeled 1 and 2 correspond to statistics only
 and statistics and systematics, respectively (similarly for dashed
 curves of the same color). The curve labeled (3) on the left has an
 additional contribution of $1\%$ systematic error on the global
 energy scale.  \label{res23} }
\end{figure}

 A great advantage of the very long baseline and multiple oscillation
 pattern in the spectrum is that the effect of systematic errors from
 flux normalization, background subtraction, and spectrum distortion
 due to nuclear effects or detector calibration can be
 small. Nevertheless, since the statistics and the size of the
 expected distortion of the spectrum are both large in the
 disappearance measurement, the final error on the precise
 determination of the parameters will most likely have significant
 contribution from systematic errors. In Figure \ref{res23} we show
 the 1 sigma resolutions that could be achieved on $\Delta m^2_{32}$
 and $\sin^2 2 \theta_{23}$.  The black lines (labeled (1)) show the
 resolutions for purely statistical errors. For the red lines (labeled
 (2)) we have included a 5\% bin-to-bin systematic uncertainty in the
 spectrum shape and a 5\% systematic uncertainty in the
 overall normalization.  These uncertainties could include modeling of
 cross sections or knowledge of the background spectra. For the
 $\Delta m^2_{32}$ resolutions we also show the expected resolution
 for an additional systematic error of 1\% on the global energy scale 
 (blue line labeled (3)).  This uncertainty for the Super
 Kamioka water Cherenkov detector is estimated to be $2.5\%$ in the
 multi-GeV region \cite{sknim}.

Although the resolution on $\Delta m^2_{32}$ will be dominated by
systematic errors for the proposed experimental arrangement, a
measurement approaching $1-2 \%$ precision can clearly be made. On the
other hand, the resolution on $\sin^2 2 \theta_{23}$ is dominated by
the statistical power at the first node. This results in a factor of
$\sim$2 better resolution with 1290 km than with 2540 km using the
same sized detector.

Running in the anti-neutrino mode with 2 MW of beam power will 
yield approximately the same spectra and resolutions on  $\Delta m^2_{32}$
and  $\sin^2 2 \theta_{23}$. By comparing the measurements with the results 
from neutrino running a test of CPT is possible.  In such a comparison
many systematic errors, such as the global energy scale,
 common to the neutrino and anti-neutrino data sets 
should cancel yielding a comparison with  errors less than $1\%$.

Finally, we remark that it is important to make precision measurements
of both $\Delta m^2_{32}$ and $\sin^2 2 \theta_{23}$ not only because
they are fundamental parameters, but also because they are needed for
interpreting the appearance ($\nu_\mu \to \nu_e$) result.  Knowledge
of both $\Delta m^2_{21}$ and $\Delta m^2_{32}$ are essential in
fitting the shape of the appearance signal to extract other
parameters. In addition, it will be very important to definitively
understand if $\sin^2 2
\theta_{23}$ is close to 1.0 or is $<1.0$. If $\sin^2 2 \theta_{23} <
1.0$ then there will be an ambiguity in $\theta_{23} \to \pi/2 -
\theta_{23}$. As we will describe below this ambiguity will affect the
interpretation of the appearance spectrum.

\section{$\nu_e$ appearance}
\begin{figure}
\vspace{5cm}
\includegraphics{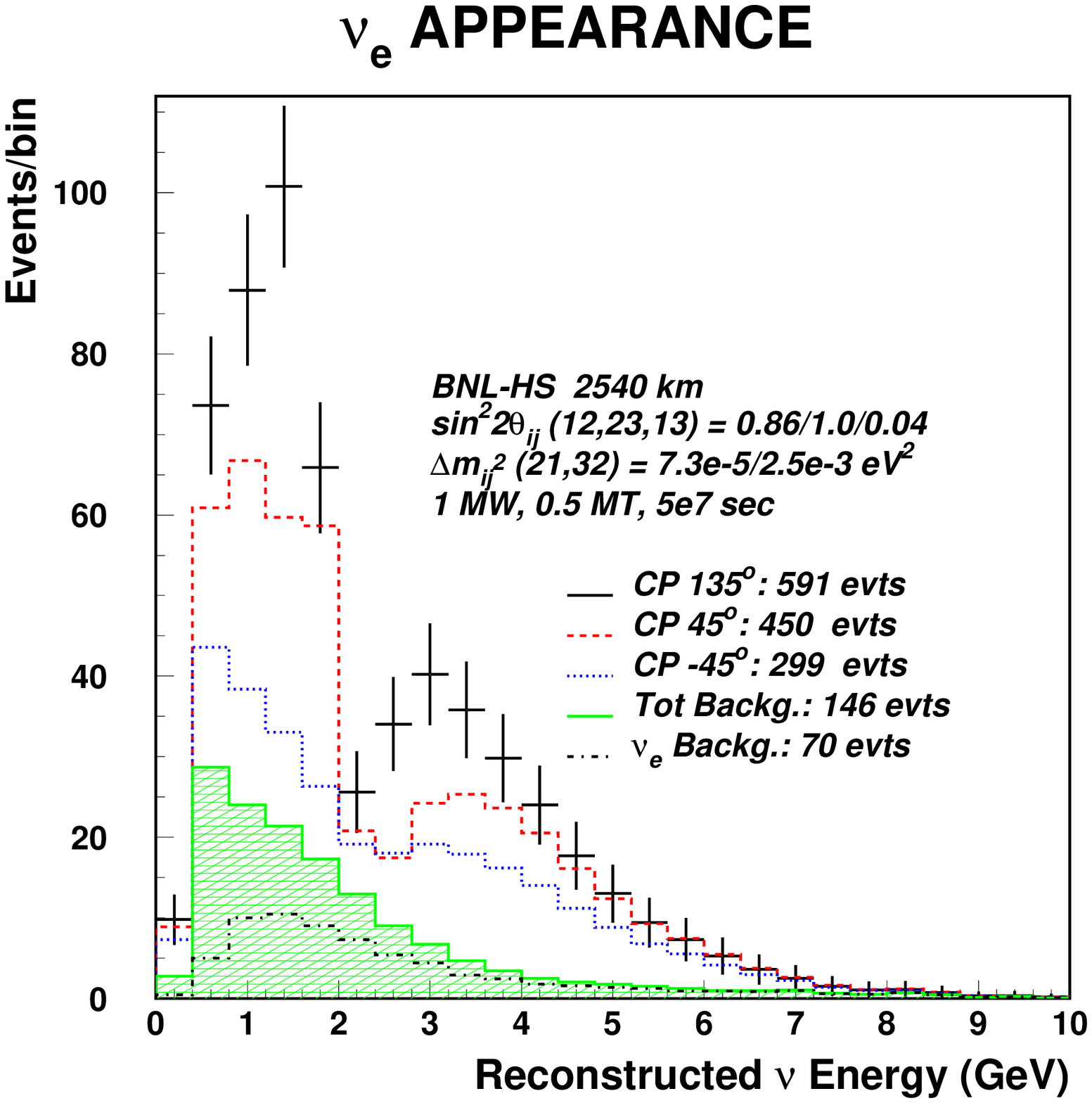}
\includegraphics{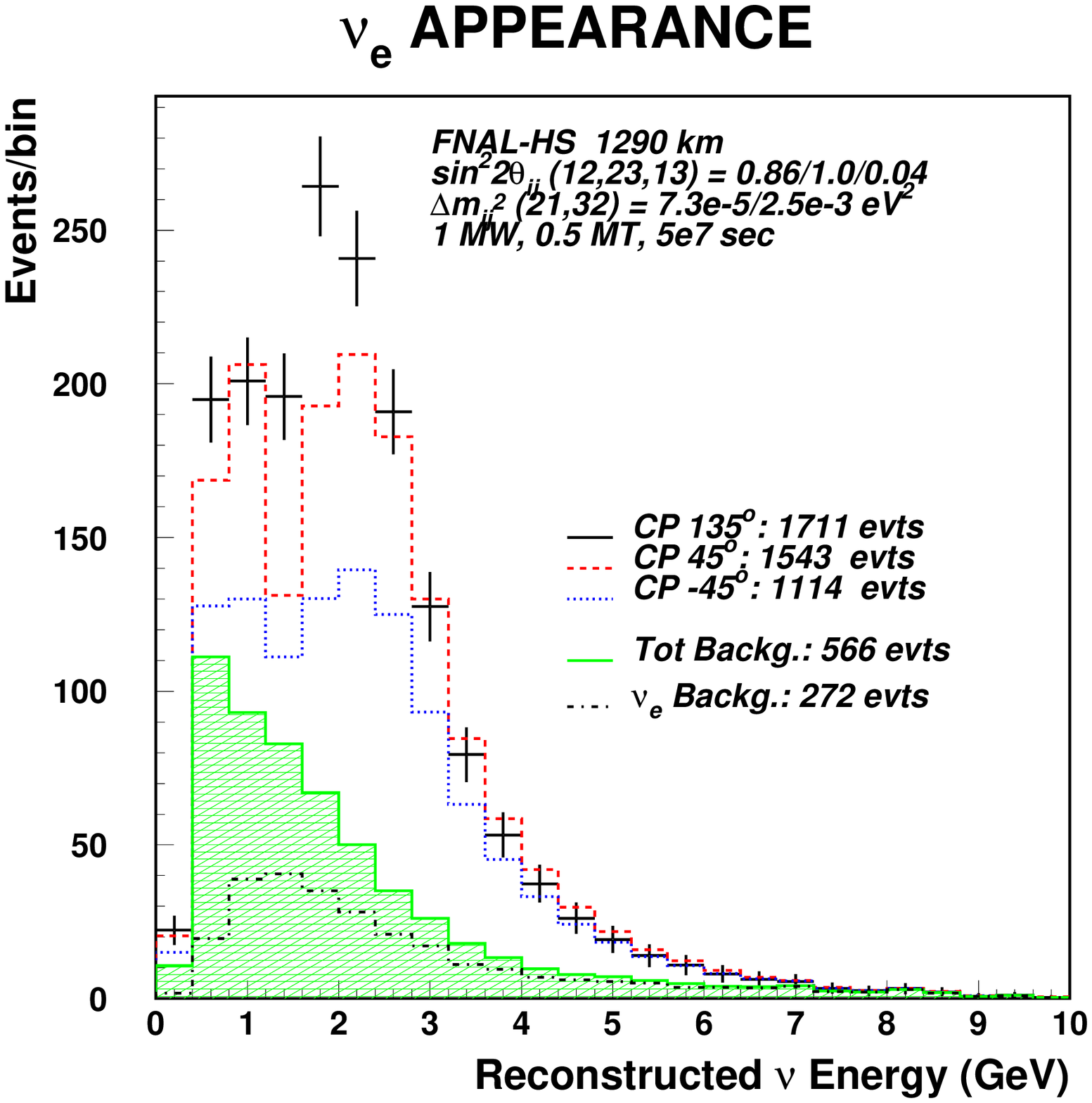}
\includegraphics{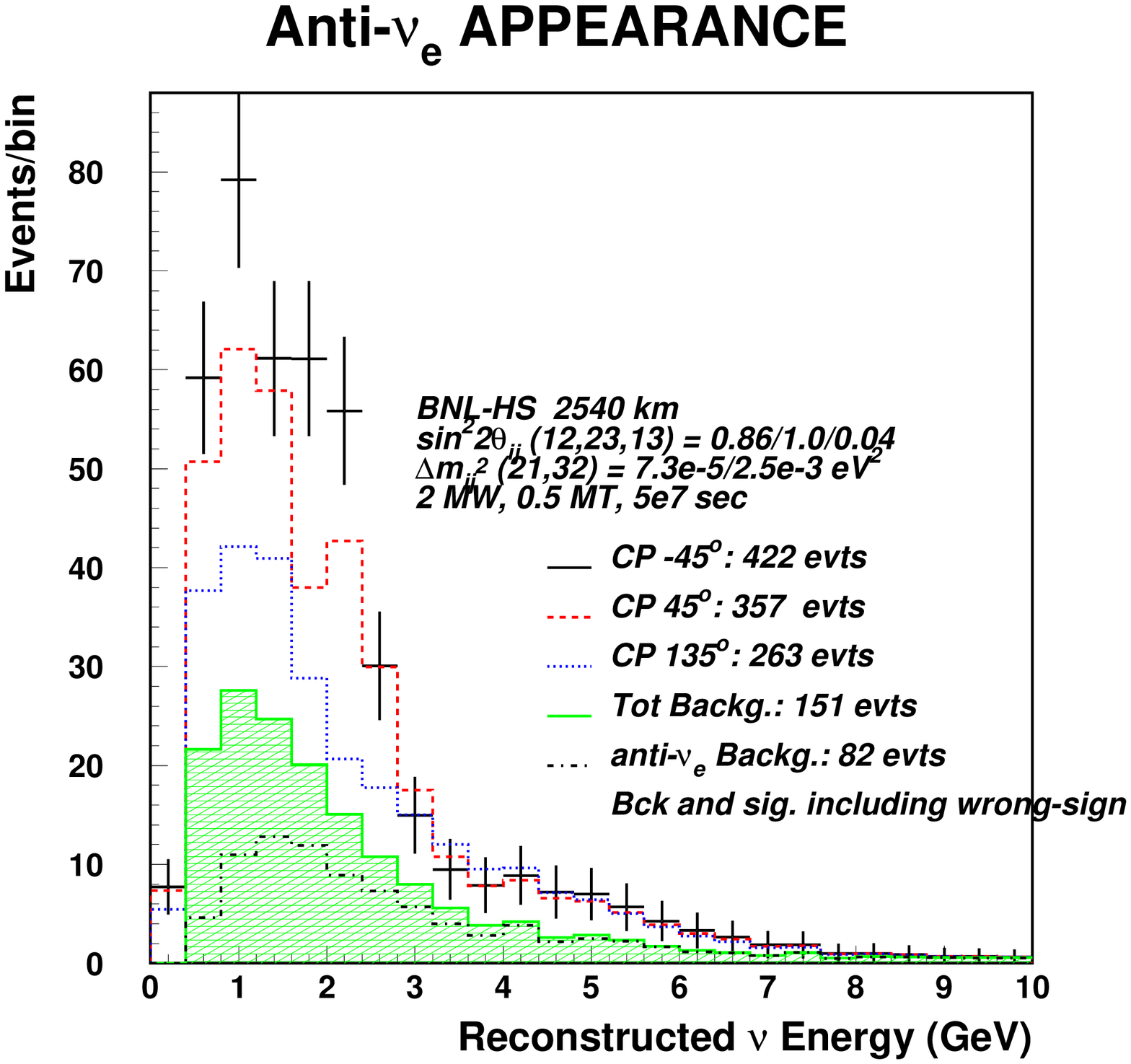}
\includegraphics{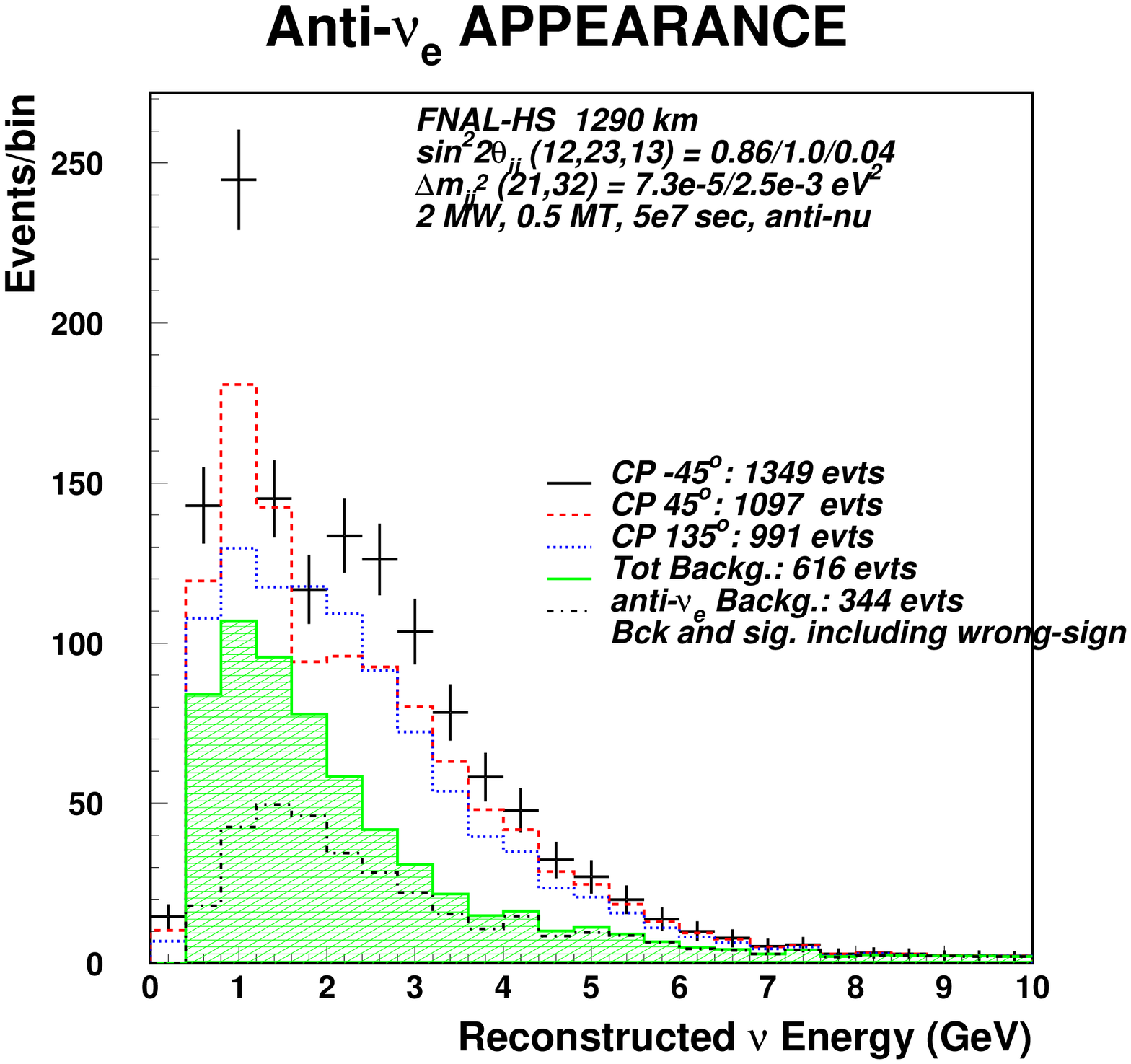}
\vspace{5.7cm}
 \caption{\it Simulation of detected electron neutrino (top plots) and
 anti-neutrino (bottom plots) spectrum (left for BNL-HS 2540km, right
 for FNAL-HS 1290 km) for 3 values of the CP parameter $\delta_{CP}$,
 $135^o$, $45^o$, and $-45^o$, including background
 contamination. Obviously, the dependence of event rate on the CP
 phase has the opposite order for neutrinos and anti-neutrinos. The
 hatched histogram shows the total background. The $\nu_e$ beam
 background is also shown. The other assumed mixing parameters and
 running conditions are shown in the figure. These spectra are for the
 regular mass hierarchy (RO).  \label{nueap} }
\end{figure}
\begin{figure}
\vspace{5cm}
\includegraphics{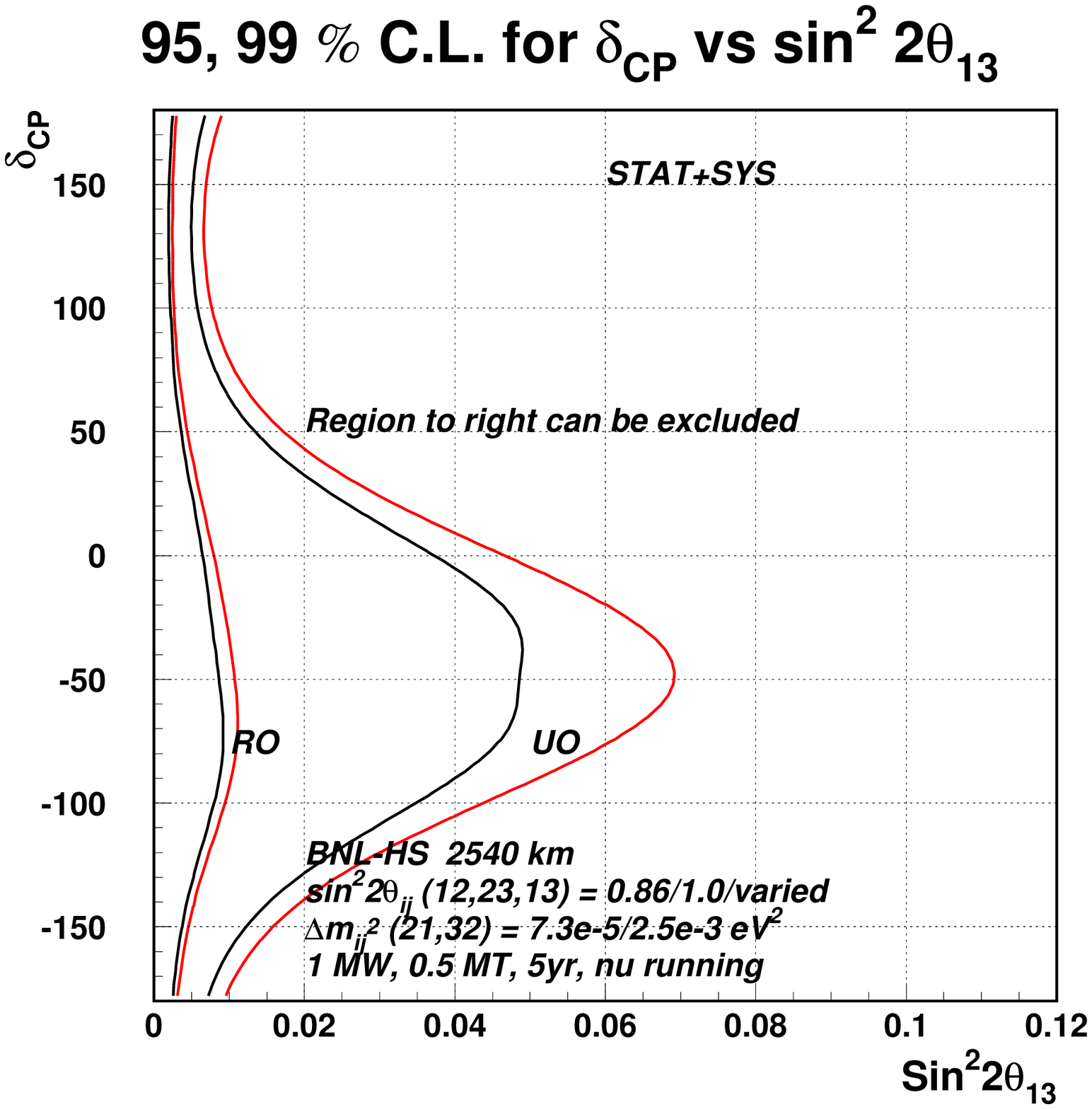}
\includegraphics{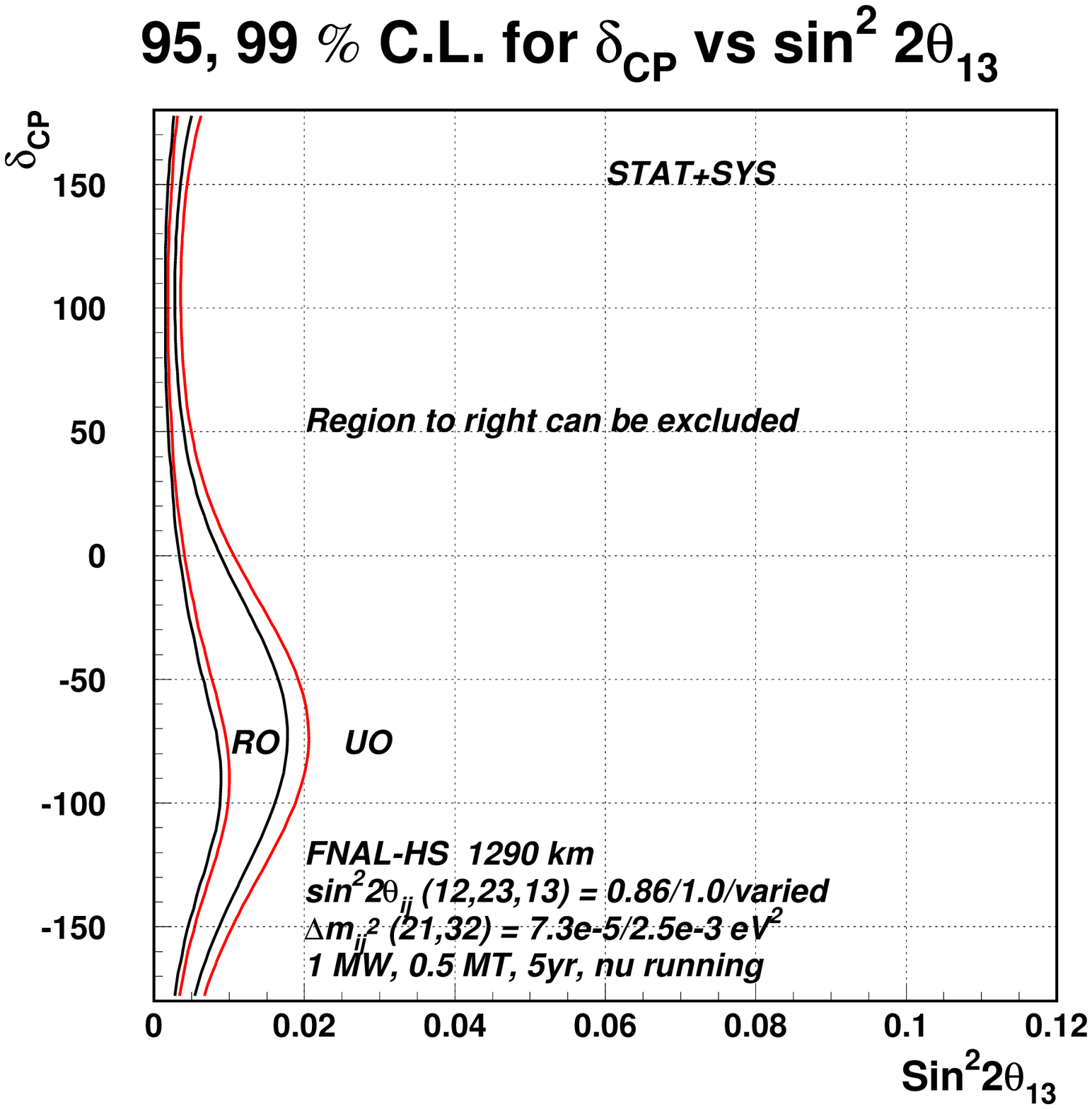} 
\includegraphics{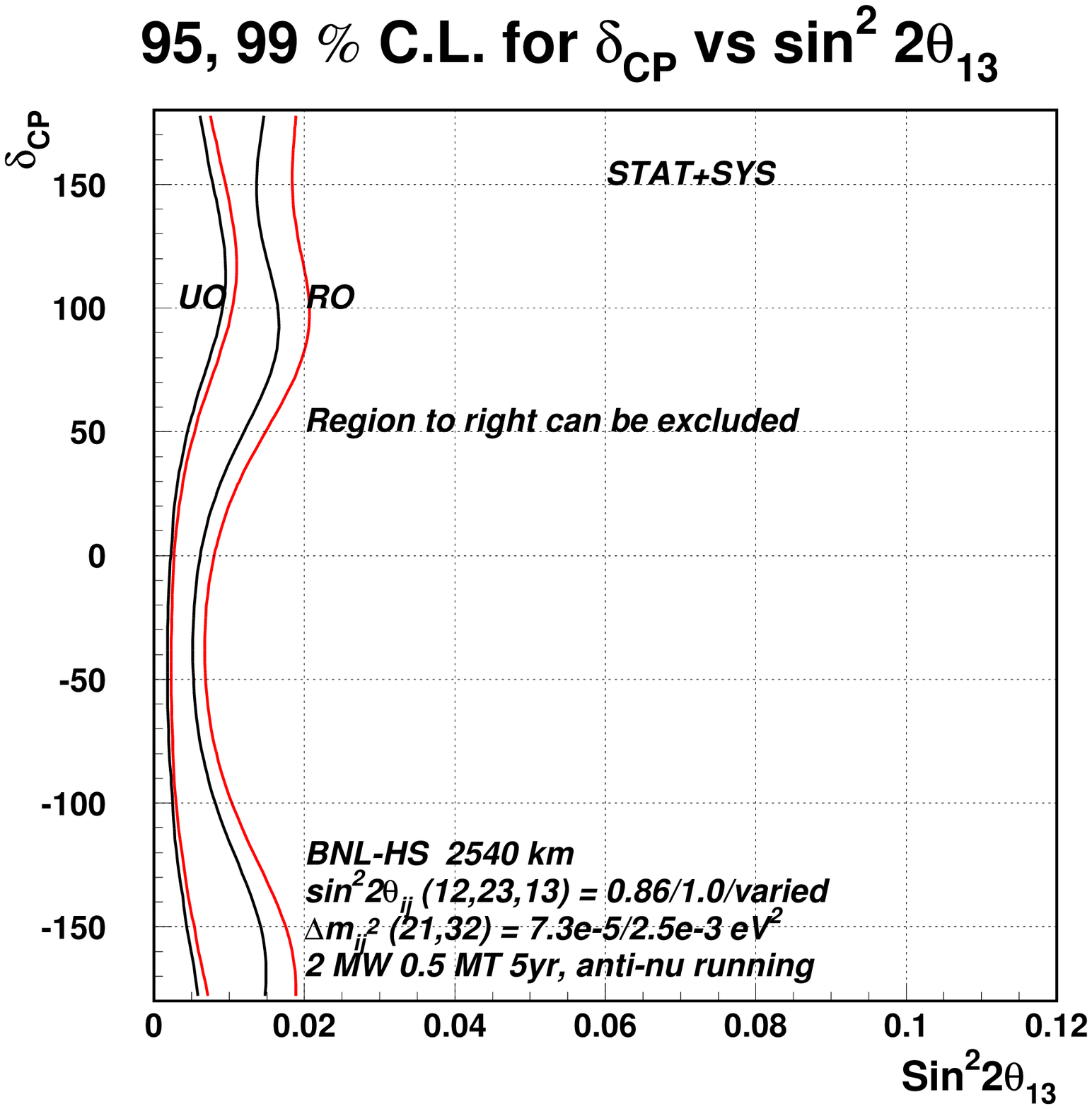}
\includegraphics{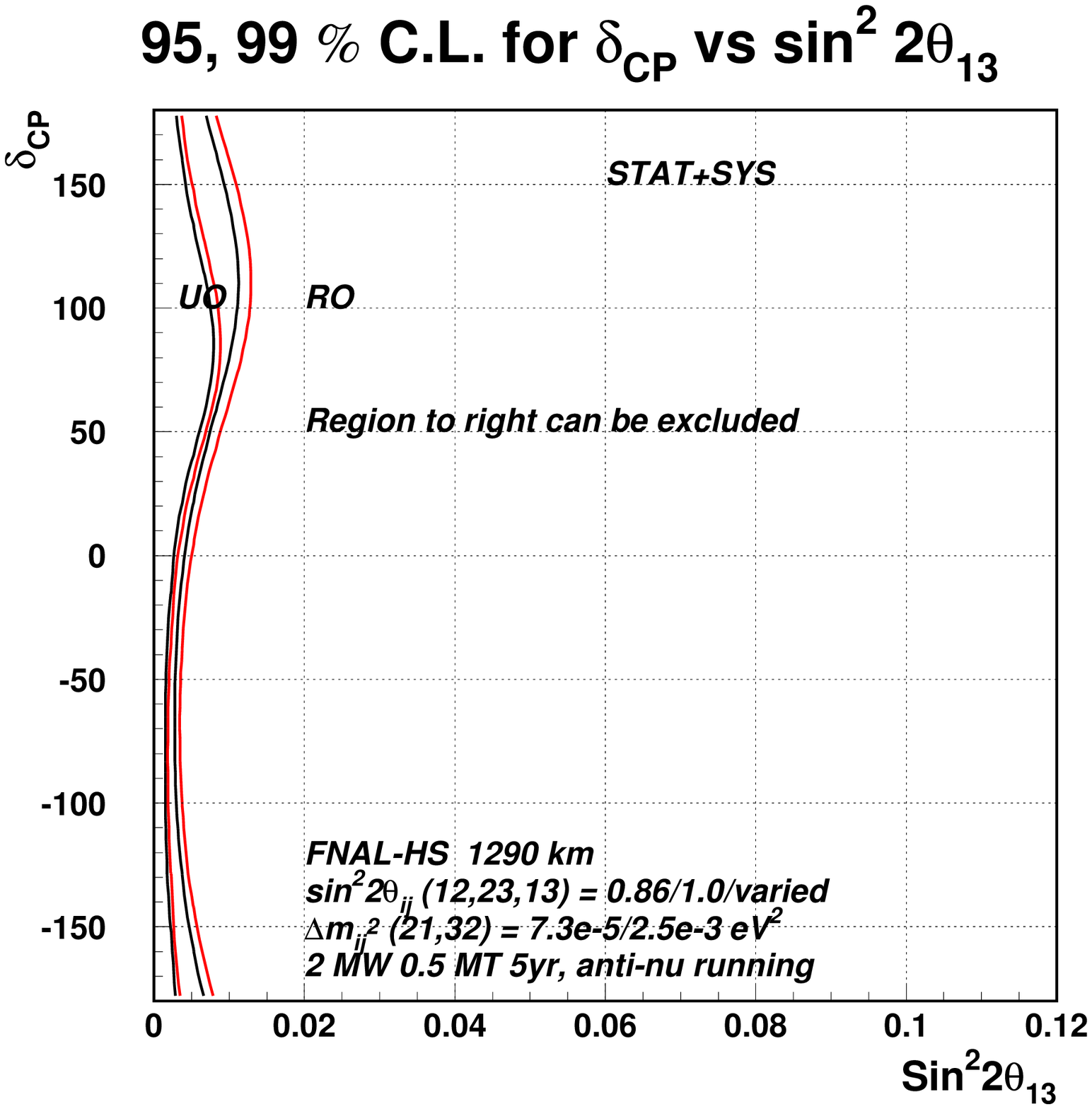}
\vspace{5.7cm}
\caption{\it
Expected limit on $\sin^2 2 \theta_{13}$ as a function of $\delta_{CP}$ for 
BNL-HS neutrino running only (top left), FNAL-HS neutrino running only (top right),
BNL-HS anti-neutrino running only (bottom left), FNAL-HS anti-neutrino running only (bottom right).
    \label{limit}}
\end{figure}
Assuming a constant matter density, 
the oscillation of $\nu_{\mu} \rightarrow \nu_e$ in the Earth 
for 3-generation mixing is described
approximately by the following equation \cite{freund}
\begin{eqnarray}
P(\nu_{\mu} \rightarrow \nu_e) &\approx&
\sin^2 \theta_{23} {\sin^2 2 \theta_{13}\over (\hat{A}-1)^2}\sin^2((\hat{A}-1)\Delta)  \nonumber\\ &&
+\alpha{\sin\delta_{CP}\cos\theta_{13}\sin 2 \theta_{12} \sin 2
\theta_{13}\sin 2 \theta_{23}\over
\hat{A}(1-\hat{A})} \sin(\Delta)\sin(\hat{A}\Delta)\sin((1-\hat{A})\Delta) \nonumber\\ &&
+\alpha{\cos\delta_{CP}\cos\theta_{13}\sin 2 \theta_{12} \sin 2
\theta_{13}\sin 2 \theta_{23}\over
\hat{A}(1-\hat{A})} \cos(\Delta)\sin(\hat{A}\Delta)\sin((1-\hat{A})\Delta) \nonumber\\ &&
+\alpha^2 {\cos^2\theta_{23}\sin^2 2 \theta_{12}\over
\hat{A}^2}\sin^2(\hat{A}\Delta) \nonumber \\
\label{qe1}
\end{eqnarray}

where $\alpha=\Delta m^2_{21}/\Delta m^2_{31}$, $\Delta = \Delta
m^2_{31} L/4E$, $\hat{A}=2 V E/\Delta m^2_{31}$,
$V=\sqrt{2} G_F n_e$. $n_e$ is the density of electrons in the Earth. 
Recall that $\Delta m^2_{31} = \Delta m^2_{32}+\Delta m^2_{21}$. 
Also notice that $\hat{A}\Delta = L G_{F} n_e/\sqrt{2}$ is sensitive 
to the sign of $\Delta m^2_{31}$. 
For anti-neutrinos, the second term in Equation \ref{qe1}  
has the opposite sign. It  is
proportional to the following CP violating quantity.

\begin{equation}
J_{CP} \equiv \sin\theta_{12} \sin\theta_{23} \sin\theta_{13} \cos\theta_{12} 
\cos\theta_{23} \cos^2\theta_{13} \sin \delta_{CP}
\label{eq2}
\end{equation}

Equation \ref{qe1}
is an  expansion in powers of $\alpha$. 
The  approximation becomes inaccurate 
 for $\Delta m^2_{32} L/4E > \pi/2$ as well as $\alpha \sim 1$. 
For the actual
results  we have used the exact numerical
calculation, accurate to all orders. 
Nevertheless, the approximate formula is
useful for understanding important features of the appearance probability:
1) the first 3 terms in the equation control the matter induced enhancement for 
regular mass ordering (RO) ($m_1< m_2< m_3$) 
or suppression for the unnatural or reversed mass ordering (UO) ($m_3< m_1< m_2$) of the oscillation 
probability above 3 GeV; 
2) the second and third terms control the sensitivity to
CP in the intermediate 1 to 3 GeV range; and 3) the last term controls
the sensitivity to $\Delta m^2_{21}$ at low energies. 

 The $\nu_e$ signal will consist of clean, single electron events (single
showering rings in a water Cherenkov detector) that result mostly 
from the
quasi-elastic reaction $\nu_e + n \rightarrow e^- + p$. The main
backgrounds will be from the electron neutrino contamination in the
beam and reactions that have a $\pi^0$ in the final state. The $\pi^0$
background will depend on how well the detector can distinguish events
with single electron induced and two photon induced 
electromagnetic  showers. Assuming the same detector performance as 
in \cite{previous} we calculate the expected electron neutrino 
and anti-neutrino spectra
shown in Figure \ref{nueap}.
These spectra were calculated for the parameters indicated in the figures 
for the regular mass ordering (RO). For the reversed mass ordering (UO)
the  anti-neutrino (neutrino) spectrum will (not) have the large matter 
enhancement at higher energies. The dependence of the total 
event rate on  the  CP phase parameter is the same for RO and UO in either 
running mode.

\subsection{$\theta_{13}$ and $\delta_{CP}$ phase}

\begin{figure}
\vspace{5cm}
\includegraphics{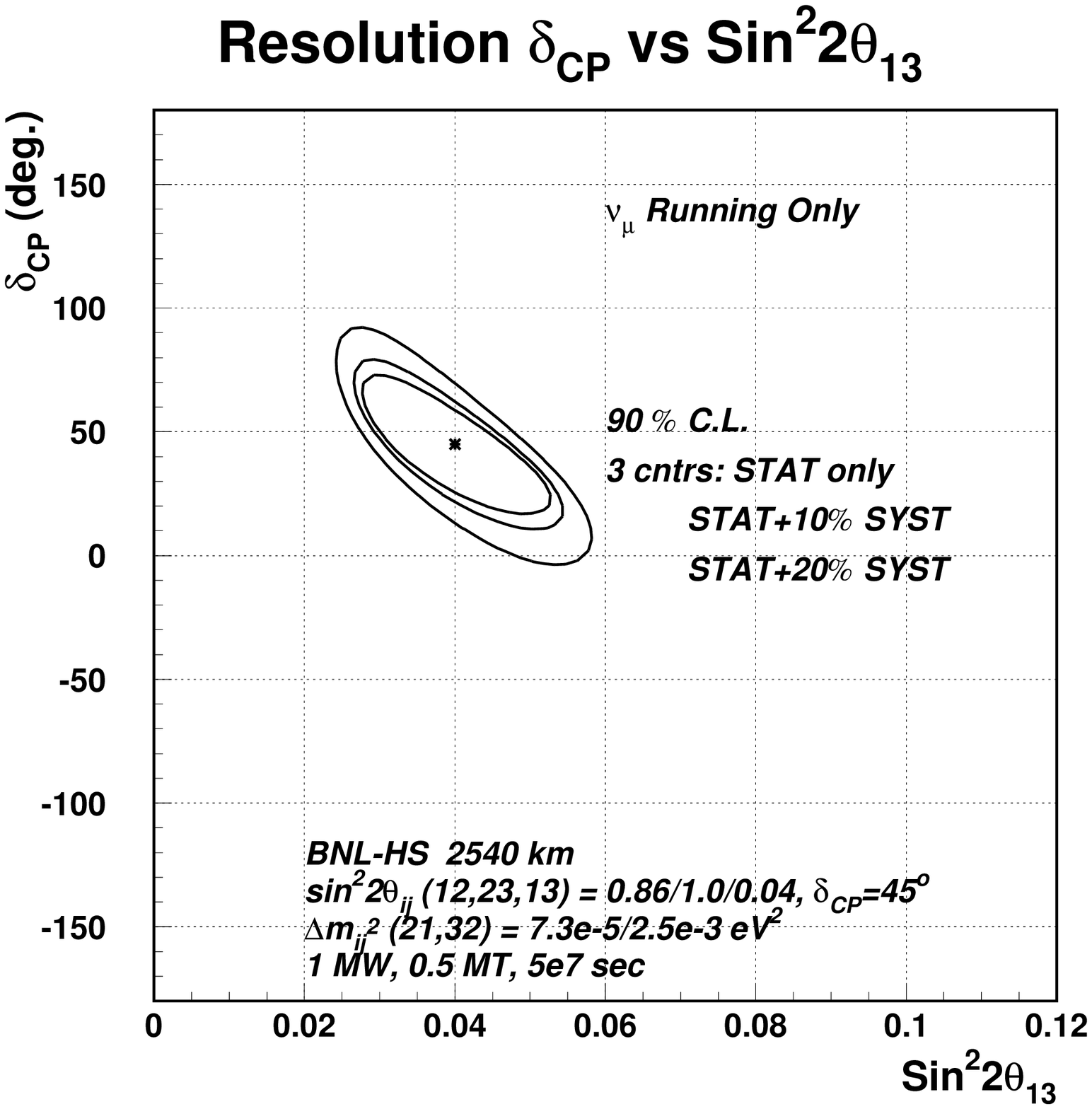}
\includegraphics{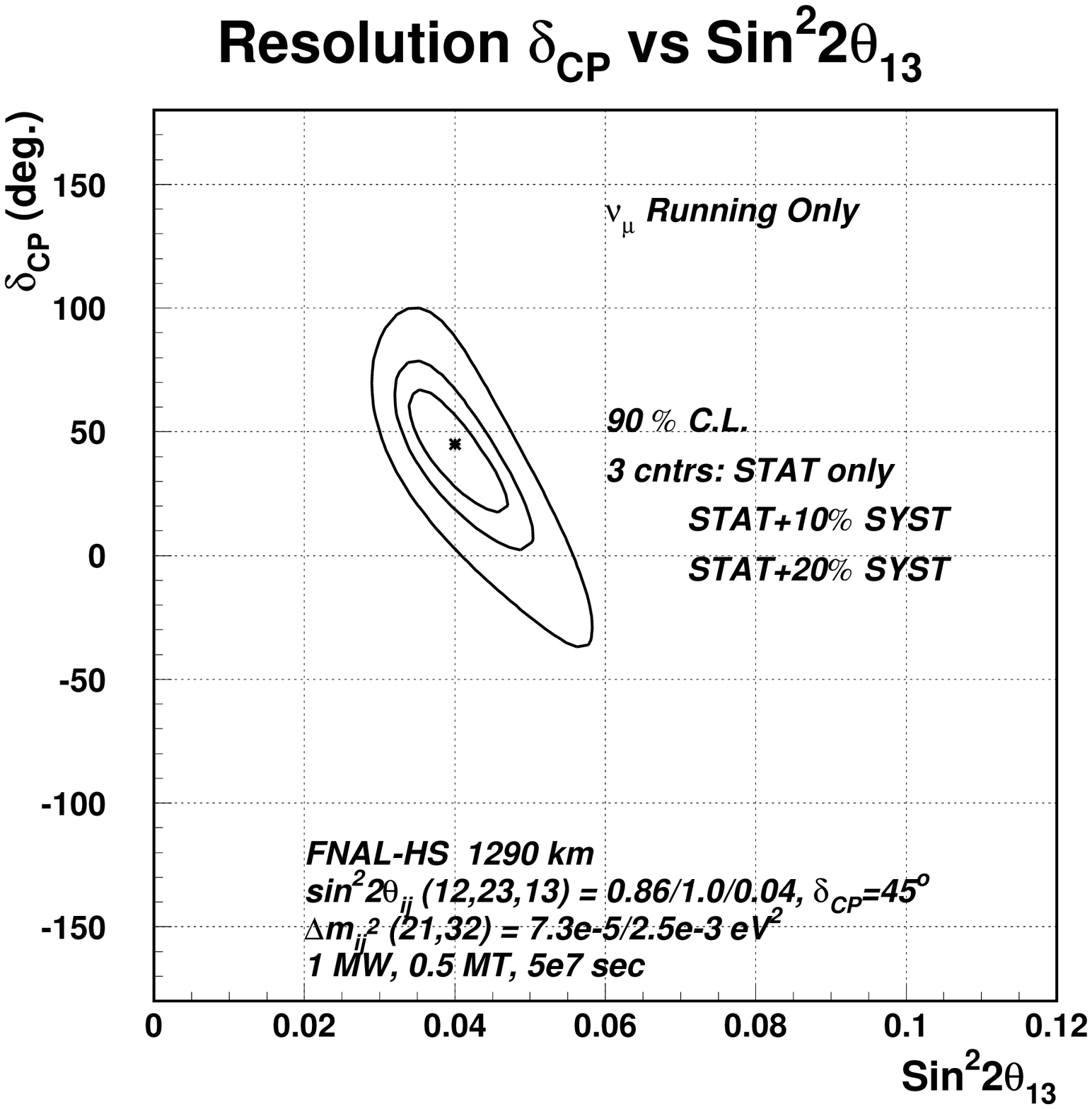}
 \caption{\it
 90\% confidence level error contours in  $\sin^2  2 \theta_{13}$ 
versus $\delta_{CP}$ for statistical and systematic errors with neutrino data alone. 
Left is for BNL-HS and right is for FNAL-HS. 
The test point used here is  
$\sin^2  2 \theta_{13}=0.04$ and $\delta_{CP}=45^o$.
 $\mdmatm = 0.0025 ~\meV^2$, and   $\mdmsol = 7.3\times 10^{-5} ~\meV^2$. The values of 
$\sin^2 2 \theta_{12}$ and $\sin^2 2 \theta_{23}$ are set to 
0.86, 1.0, respectively. 
    \label{cpmea} }
\end{figure}

\begin{figure}
\vspace{5cm}
\includegraphics{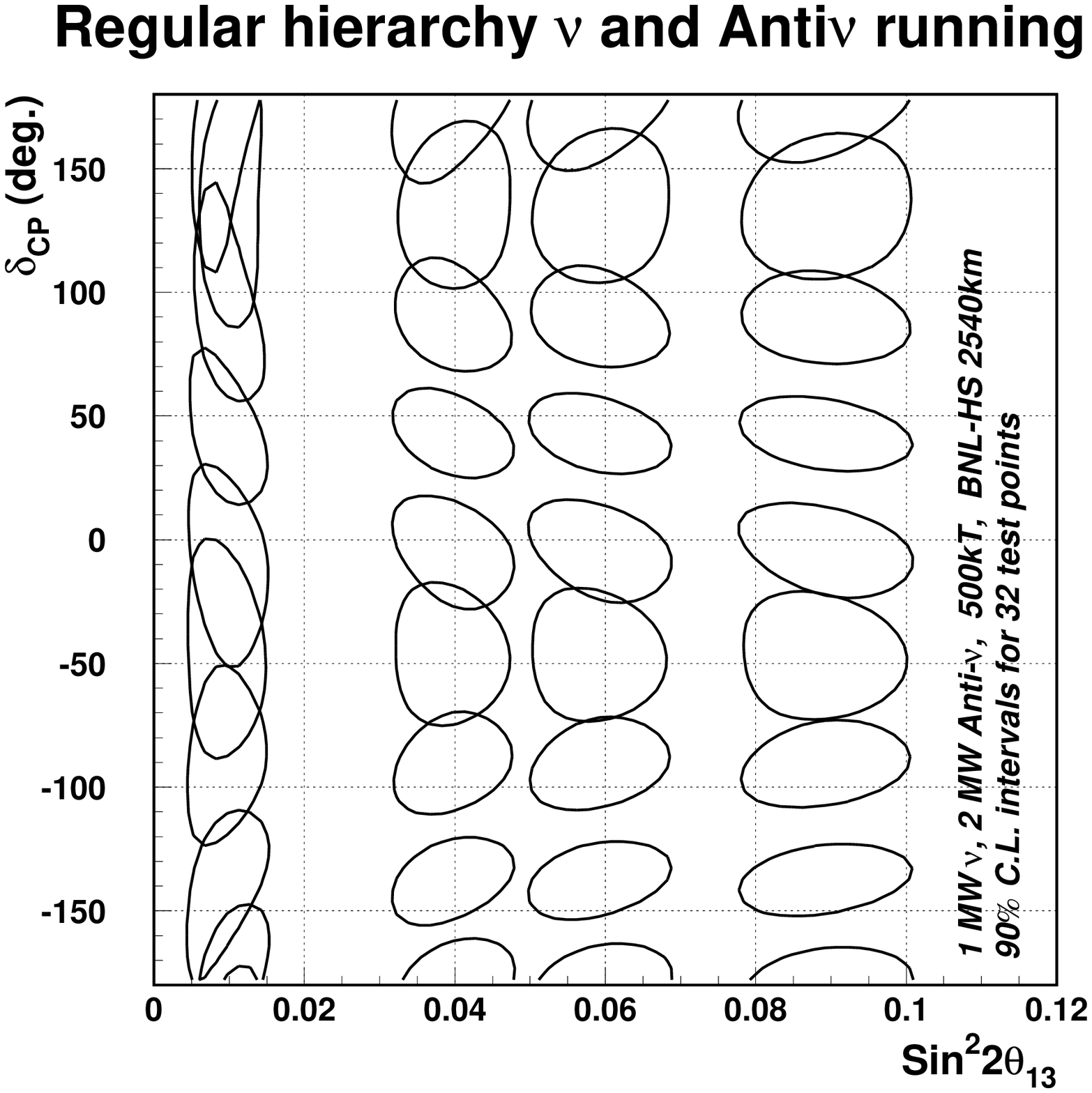}
\includegraphics{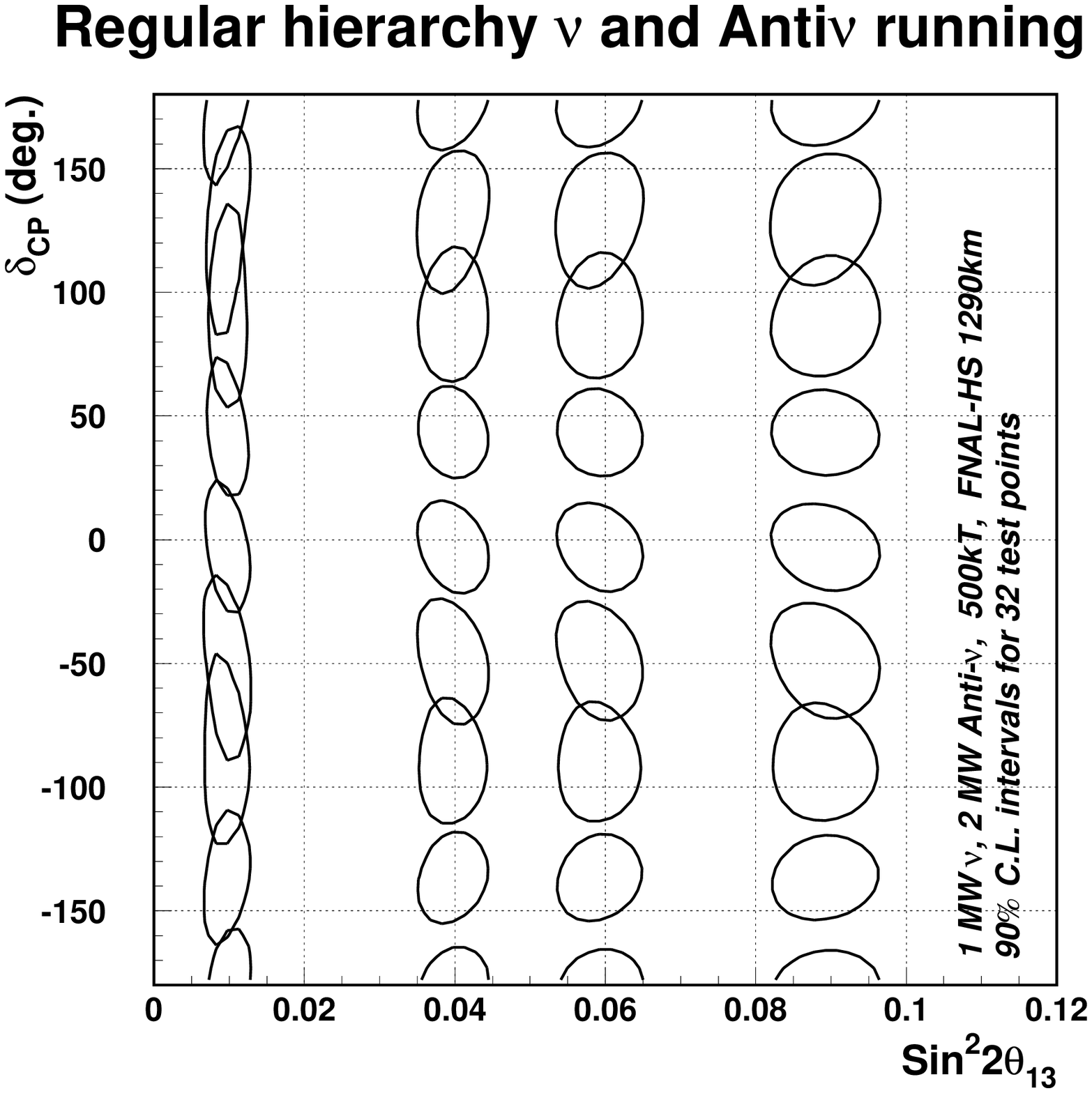}
 \caption{\it
 90\% confidence level error contours in  $\sin^2  2 \theta_{13}$ 
versus $\delta_{CP}$ for statistical and systematic errors for 32 test points.
This simulation is for combining both  neutrino and anti-neutrino data.  
Left is for BNL-HS and right is for FNAL-HS. 
We assume 10\% systematic errors for this plot. 
    \label{nuanucp} }
\end{figure}

If there is no excess of electron events observed then we can set a
limit on the value of $\sin^2 2 \theta_{13}$ as a function of
$\delta_{CP}$.  Such 95 and 99\% C.L. sensitivity limits are shown in
Figure \ref{limit}. These set of plots illustrate various
considerations that must be evaluated for the very long baseline
project.  After running initially in the neutrino mode with 1 MW of
beam power, if an excess signal is found then a measurement of
$\delta_{CP}$ versus $\sin^2 2 \theta_{13}$ can be made as shown in
Figure \ref{cpmea}, at the same time the mass hierarchy is determined
from the strength of the signal in the higher energy region.  If there
is no signal in the neutrino mode then either $\theta_{13}$ is too
small for the regular mass hierarchy (RO) or the mass hierarchy is reversed
(UO) and parameters are in the ``unlucky'' region
($-140^o<\delta_{CP}<30^o$). For the shorter baseline of 1290 km, the
$\theta_{13}$ sensitivity for the reversed hierarchy is not reduced as
much as for 2540 km because both the CP-sensitivity and the matter
effect are weaker. Although this yields a better limit for $\sin^2 2
\theta_{13}$ in the absence of signal, 
it affects the precision on $\delta_{CP}$ and the determination of 
the mass hierarchy.

If there is no signal in the neutrino mode, we will  run in the
anti-neutrino mode to cover the ``unlucky'' parameter space for the
appearance signal. A
combination of neutrino and anti-neutrino running will yield a
stringent limit approaching $\sin^2 2 \theta_{13} \sim 0.003$
independent of the value of $\delta_{CP}$.  The simulation results
shown here include wrong sign contamination in both the background and
signal for anti-neutrinos. Interestingly,  since more than 20\% of the event rate in
the anti-neutrino case  actually arises from
the neutrino contamination, the $\sin^2 2 \theta_{13}$ limit in the
anti-neutrino case exhibits  less dependence on $\delta_{CP}$ and
the mass hierarchy.
If there is a signal in the neutrino mode, we will get the first
measurement of $\delta_{CP}$ from neutrino data alone in the
3-generation model, but it will still be important to run in the
anti-neutrino mode for better precision, 
over-constraints on the 3-generation model,
and search for possible new physics
either in the mixing  or in the interactions of neutrinos. 

In Figure \ref{cpmea} we show the 90\% confidence level
interval in the $\delta_{CP}$ versus $\sin^2 2 \theta_{13}$ plane from
neutrino running alone for the two baselines.  We have chosen the
point $\delta_{CP}=45^o$ and $\sin^2 2 \theta_{13}=0.04$ as an
example.  At this test point for the regular mass hierarchy, the
resolution on $\delta_{CP}$ is $\sim \pm 20^o$. The mass hierarchy is
also resolved at $>5$ sigma because of the large enhancement of the
spectrum at higher energies.  As we pointed out in the introduction,
the resolution on the CP phase is approximately independent of the
baseline. The major difference between the 1290 and 2540 km baselines
is that the shorter baseline has higher correlation between the
parameters, $\delta_{CP}$ and $\sin^2 2 \theta_{13}$, has better
resolution on $\sin^2 2 \theta_{13}$, and has worse sensitivity to
systematic errors on the background and the spectrum shape. If 
the systematic errors exceed 10\%, the shorter
baseline will most likely have worse performance for measuring the CP
parameter. 

The sensitivity to systematic errors and the
dependence on the mass hierarchy can be relieved by using data from
both neutrino and anti-neutrino running. 
Figure \ref{nuanucp} shows the 90\% confidence level interval for 32
test points in the $\delta_{CP}$ and $\sin^2 2 \theta_{13}$ plane
after both neutrino and anti-neutrino data.  A
number of observations can be made: Figure \ref{nuanucp} is for the
regular mass hierarchy. The plot for the reversed mass hierarchy is
similar. After both neutrino and anti-neutrino data the hierarchy will
be resolved to more than 10 sigma (somewhat
 less significance for the shorter baseline) 
  for $\sin^2 2 \theta_{13}$ as small
as 0.01. The resolution on $\delta_{CP}$ is seen to be approximately
independent of $\sin^2 2 \theta_{13}$ for $\sin^2 2
\theta_{13}>0.01$. When $\sin^2 2 \theta_{13}$ is so small that the
background becomes dominant, the $\delta_{CP}$ resolution becomes
poor.  The resolution on $\delta_{CP}$ is seen to be approximately the
same for 2540 and 1290 km, except for small $\sin^2 2 \theta_{13}$
where large statistics at 1290 km are seen to overcome the
background. The resolution on $\sin^2 2 \theta_{13}$ is, however,
better for the shorter baseline because the sensitivity comes from the
first node of oscillations which has much higher statistics at the
shorter baseline.

\subsection{Correlations with other parameters}

The measurement of $\delta_{CP}$ using a wide band beam and multiple
oscillation nodes is largely free of ambiguities and correlations
\cite{barger}. The $\delta_{CP} \to \pi-\delta_{CP}$ ambiguity is
resolved by the detection of multiple nodes including the effects of
the $\cos \delta_{CP}$ term. The mass hierarchy is resolved because
it has a strong energy dependence obvious in the shape of the
spectrum. 

The remaining main sources of correlations are the
uncertainty on $\Delta m^2_{21}$ and $\sin^2 2 \theta_{23}$.
The CP terms in Equation \ref{qe1} are linear in $\Delta m^2_{21}$, therefore
the systematic uncertainty on the event rate at the second oscillation
maximum will be $<10\%$, which is the uncertainty on $\Delta m^2_{21}$
from solar neutrino and KAMLAND experiments.  As discussed above, this
level of uncertainty will not affect the CP measurement for the longer
baseline of 2540 km, but could be important for the shorter baseline of 1290 km. 

An examination of Equation \ref{qe1} shows that the knowledge of
$\theta_{23}$ affects the first ($\Delta m^2_{31}$ dominated) and the
last ($\Delta m^2_{21}$ dominated) terms as $\sin^2 \theta_{23}$ and
$\cos^2 \theta_{23}$, respectively.  The first term is responsible for
the matter enhanced (or suppressed) 
appearance at high energies and the last term is responsible for
appearance at  low energies. Current knowledge of $\theta_{23}$ from 
atmospheric neutrinos  \cite{sk}  is rather poor:
$35 <\theta_{23} <55^o$.  A precise determination of $\sin^2 2
\theta_{23}$ using the muon disappearance spectrum is, therefore,  
essential for
proper interpretation of the appearance signal.  A 1\% determination
of $\sin^2 2 \theta_{23}$ (Figure \ref{res23}) leads to an uncertainty
of $\sim 10\%$ on the appearance event rates if $\theta_{23} \sim
45^o$ and $\sim 2\%$ if $\theta_{23} \sim 35^o$. If $\theta_{23} \sim 35^o$
then there is also the additional ambiguity of $\theta_{23} \to
\pi/2-\theta_{23}$. Because of the strong energy dependence at low and
high energies the ambiguity as well as the uncertainty should not
affect the interpretation of the neutrino data in the case of the
longer 2540km baseline.  Uncertainties on both $\Delta m^2_{21}$ and
$\theta_{23}$ affect the neutrino and anti-neutrino appearance spectra
in the same manner, therefore after both data sets are acquired these
systematic errors are expected to have little effect on establishing
CP violation in neutrinos, but may affect the determination of
parameters in the case of the shorter baseline.

\begin{figure}
\vspace{5cm}
\includegraphics{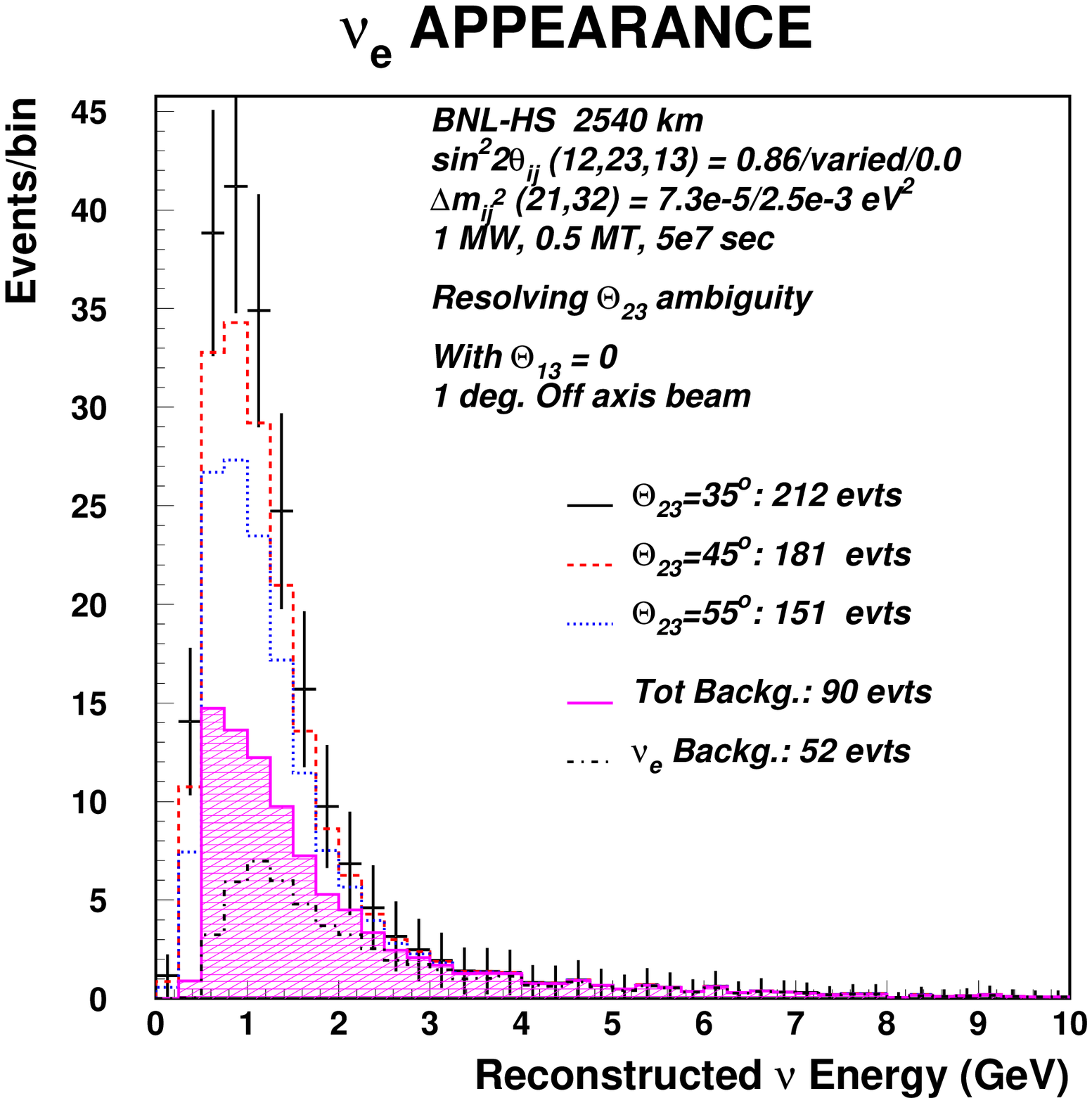}
\includegraphics{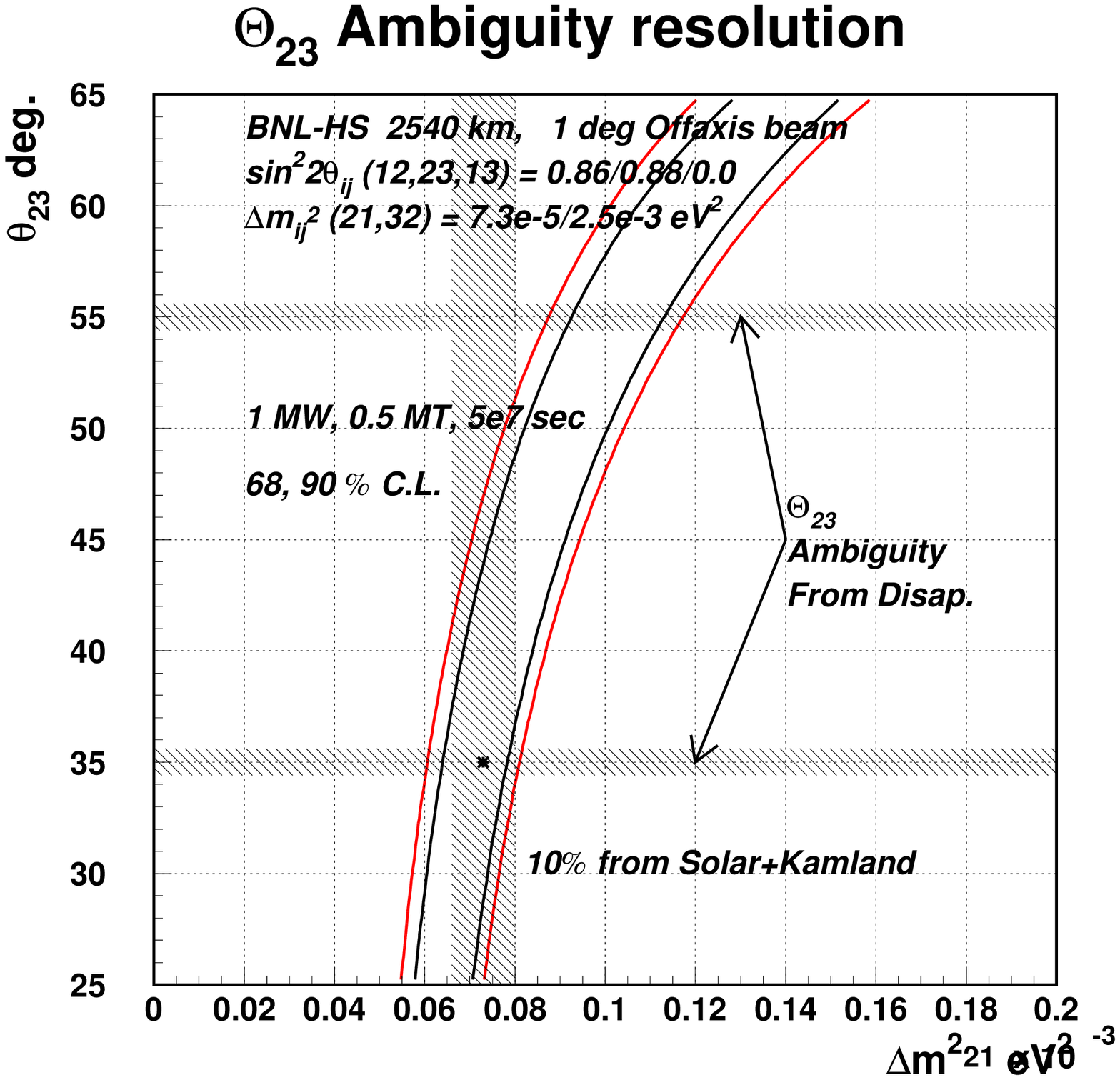}
 \caption{\it
Expected  spectrum of electron neutrinos (left)  for $\theta_{13}=0$ and other assumed 
parameters indicated in the figure. 
The right hand side shows the resolution of the $\theta_{23} \to \pi/2-\theta_{23}$ ambiguity 
using the measurement of $\sin^2 2 \theta_{23}$ from disappearance and assuming a 
10\% measurement of $\Delta m^2_{21}$ from KAMLAND.  The area between the 
curves is allowed by the appearance spectrum (left) for $\theta_{23}=35^o$.
    \label{th23amb} }
\end{figure}

It is important to understand the physics case for the super-beam if
$\sin^2 2 \theta_{13}$ is so small that the background prevents us
from detecting a signal. In this case, both the mass hierarchy through
the matter effect and the CP phase measurement are not accessible for
any baseline.  However, the $\nu_\mu \to \nu_e$ conversion signature
still could be accessible for the longer baseline of 2540 km because
of the last term in Equation \ref{qe1}.  This term depends on the
``solar'' $\Delta m^2_{21}$ as well as $\sin^2 2 \theta_{12}$ and
$\cos^2 \theta_{23}$. For the current value of the solar parameters
$\sim 100$ events could be expected over a similar background.  This
is shown in Figure \ref{th23amb} where we have used a 1 degree off-axis
neutrino spectrum to reduce the background level at low energies.  For
this calculation we have used $\sin^2 2 \theta_{23}=1.0$ 
and $\sin^2 2 \theta_{23}=0.88$ as  test
points.  We assume that $\sin^2 2 \theta_{23}$ will be measured with $\sim$1.5\%
precision in disappearance. 
In the case of $\sin^2 2 \theta_{23}=0.88$, we are 
 lead to an ambiguity in $\theta_{23}$ of $35^o\pm 0.6^o \to
55^o\pm 0.6^o$.  This ambiguity is clearly distinguished at several 
sigma in the case of
the 2540 km baseline as shown in the right hand side of 
Figure \ref{th23amb}. The ambiguity 
resolution is accomplished by comparing the result of appearance with 
the result of $\bar\nu_e$ disappearance from solar and KAMLAND 
measurements. For Figure \ref{th23amb}
we assume that $\Delta m^2_{21}$ will be determined to $\sim 10\%$. 
This comparison of appearance and disappearance experiment could also 
be important for uncovering new physics in this sector.

\section{Conclusion}
We have studied various possible measurements using a powerful
neutrino beam, using a MW-class proton source located either at BNL or
FNAL, to a large capable detector with fiducial mass in excess of
100kT over a distance $\sim 2000$km. For our study here, we chose the
distances of 1290 and 2540km because they correspond to the distances
from FNAL and BNL to Homestake in South Dakota, one of the possible
sites for a large detector.  Nevertheless, our results are applicable
to any other site in the U.S. at a comparable distance from an
accelerator laboratory.  Qualitatively, this project is motivated by
the need to perform an experiment that is sensitive to both the
atmospheric ($\Delta m^2_{32}\sim 0.0025 eV^2$ ) and the solar ($\Delta
m^2_{21}\sim 8\times 10^{-5} eV^2$ ) oscillation scales and to obtain
an oscillatory pattern in the energy spectrum of muon neutrinos.  The
detector requirements for such an experiment -- both in size and
performance -- are well-matched to other important goals in particles
physics, such as detection of proton decay and astrophysical neutrinos. 
Therefore the potential physics impact is
very broad for particle and astrophysics. 

In this paper we have shown that very precise measurements of $\Delta
m^2_{32}$ and $\sin^2 2 \theta_{23}$ can be made using the observation
of the oscillatory spectrum of muon neutrinos at either 1290 or 2540
km.  For these precise measurements the shorter baseline has an
advantage because of the increased statistical power, however it is
very likely that the measurements will be systematics dominated to
about 1\% for either distance.  We have also shown that very good
bounds on $\sin^2 2 \theta_{13}$ can be obtained from both baselines
using the appearance of electron neutrinos.  The electron event rate at
shorter baseline has smaller matter effect and smaller dependence on
the CP phase. Therefore, the $\theta_{13}$ bound using the neutrino
data alone from the shorter baseline will have less dependence on the
CP phase and the mass hierarchy. When both neutrino and
anti-neutrino data are combined the $\delta_{CP}$ and mass hierarchy
dependence is eliminated for both baselines, and the $\theta_{13}$
bound from either baseline will  likely be dominated by the
knowledge of backgrounds. The limit on $\sin^2 2 \theta_{13}$ could
reach  $\sim 0.003$ if total number of background
events can be controlled to about twice the expectation from
the electron neutrino contamination in the beam ($\sim 0.7\%$).

If a signal is found for electron appearance then the value of the CP
phase can be determined from the shape of the spectrum using neutrino
data alone for either baseline.  A more precise measurement of the CP
phase and further constraint on the 3-generation model can be made by
additional running in the anti-neutrino mode.  There are some
advantages for having the longer 2540 km baseline for the CP
measurement. The matter effect is much larger and therefore the mass
hierarchy can be resolved with greater confidence. The effect of
$\delta_{CP}$ on the spectrum is also much larger for the longer
baseline. This allows extraction of the parameter $\delta_{CP}$
without relying on very precise determination of the spectrum
shape. The systematics of the spectrum shape are dependent not only on
the knowledge of the beam, but also on  other neutrino parameters
such as $\Delta m^2_{31}$, $\theta_{23}$, $\Delta m^2_{21}$, and
$\theta_{12}$.  These parameters must be obtained from solar and
reactor experiments, and from the muon neutrino 
disappearance analysis. A 10\%
systematic uncertainty on the backgrounds and the shape of the
spectrum is tolerable for the 2540 km baseline, whereas the
uncertainty needs to be smaller for the shorter baseline
experiment. In addition, the longer baseline allows detection of the
appearance of electron neutrinos even if $\theta_{13}$ is too small,
through the effect of $\Delta m^2_{21}$ alone.  This observation can
also help separate the $\theta_{23} \to \pi/2 -\theta_{23}$ ambiguity
if needed.

Despite the small, but significant differences between the two
possible baselines, we conclude that an experiment using a beam from
either FNAL or BNL to a large next generation multipurpose detector is
very important for particle physics and could lead to major advances
in our understanding of neutrino phenomena. It is important to
recognize that the detector meant for such an experiment needs to be
highly capable in terms of pattern recognition and energy
resolution. If such a detector is located in a deep low background
environment, it has broad applications in searching for nucleon decay
and astrophysical neutrino sources.  There are many advantages
if both beams  can be built and sent to the same detector. The
correlation between parameters and the size of the matter
effect are different for the two baselines. It is possible that by
combining the results from the two baselines all dependence on
external parameters could be eliminated, and the neutrino sector much
better constrained. The requirement on total running time could also
be reduced.

This work was
supported by DOE grant DE-AC02-98CH10886. I also want to thank the Aspen Center for 
Physics where much of the writing of this paper took place.


\begin{thebibliography}{99}



\bibitem{previous} 
``Very Long Baseline Neutrino Oscillation Experiment for Precise 
Measurements of Mixing Parameters and CP Violating Effects'',
M. V. Diwan, {\em et al.}, {\bf PRD 68} (2003) 012002.


\bibitem{ref1} PDG, Phys. Rev. {\bf D66}, 010001 (2002), p. 281.

\bibitem{ref3} W. Marciano, hep-ph/0108181, 22 Aug. 2001.

\bibitem{ref4} Stephen Parke, Talk in the HQL04 conference, 
Puerto Rico, June 1-June5, 2004. 

\bibitem{k2k}
S. H. Ahn et al., Phys. Lett. {\bf B 511} 178 (2001).

\bibitem{minos} Numi MINOS project at 
Fermi National Accelerator Laboratory, http:/www-numi.fnal.gov/ 


\bibitem{CNGS}   The CERN Neutrino beam to Gran Sasso (Conceptual Technical Design),
 Ed. K. Elsener, CERN 98-02, INFN/AE-98/05, http://proj-cngs.web.cern.ch/proj-cngs/



\bibitem{offaxis} 
The JHF-Kamioka neutrino project, Y. Itow et al., 
arXiv:hep-ex/0106019, June 2001.

\bibitem{numioff}
Letter of Intent to build an Off-axis Detector to study numu to nue
oscillations with the NuMI Neutrino Beam, D. Ayres {\it et al.},
hep-ex/0210005.



\bibitem{sk} Y.~Fukuda et al., Phys. Rev. Lett. {\bf 81}, 1562 (1998);
S. Fukuda et al., Phys. Rev. Lett. {\bf 86} 5656, 2001;  
E.W. Beier, Phys. Lett. {\bf B283}, 446 (1992); 
T. Kajita and Y. Totsuka, Rev. Mod. Phys. {\bf 73}, 85 (2001).

\bibitem{kamland} K. Eguchi, {\it et. al}, Phys. Rev. Lett. {\bf 90},
  021802 (2003). hep-ex/0212021.

\bibitem{sno} Q. R. Ahmad et al., Phys. Rev. Lett. {\bf 87} 071301 (2001). 
S. Fukuda et al., Phys. Rev. Lett., {\bf 86} 5651 (2001). 

\bibitem{agsup} J. Alessi {\it et al.}, AGS Super Neutrino Beam 
Facility, Accelerator and Target system Design, BNL-71228-2003-IR. 
April 15, 2003.  http://nwg.phy.bnl.gov/

\bibitem{NUSEL}
Neutrinos and Beyond: New Windows on Nature,
Neutrino Facilities Assesment Committee, National Research Council,
(2003), ISBN-0-309-087 16-3, http://www.nap.edu/catalog/10583.html.


\bibitem{foster} R. Alber, et al.,  Accelerator Proton Driver Study Group
FNAL-TM-2136, FNAL-TM-2169. 
http://www-bd.fnal.gov/pdriver/




\bibitem{3m} Megaton Modular Multi-Purpose Neutrino detector, 
3M collaboration, http://www.hep.upenn.edu/Homestake

\bibitem{uno} Physics Potential and Feasibility of UNO, 
UNO collaboration, June 2001, Stony Brook University, SBHEP01-03. 


\bibitem{icarus} 
D.B.~Cline, F. ~Segiampietri, J.G.~Learned, K.T.~McDonald,
{\sl LANNDD, A Massive Liquid Argon Detector for Proton Decay,
Supernova and Solar neutrino Studies, and a Neutrino Factory 
Detector}{May 24, 2001}, astro-phy/0105442; 
F. Arneodo et al., Nucl. Instrum. Meth. {\bf A471} 272-275 (2000).

\bibitem{e889} The ``off-axis'' neutrino beam was first proposed by the
 E889 Collaboration, Physics Design Report, 
BNL No. 52459, April, 1995. http://minos.phy.bnl.gov/nwg/papers/E889.


\bibitem{sknim}  S. Fukuda et al., Nucl. Instrm. Meth. A {\bf 501}, 
(2003) 418-462. 

\bibitem{ypsilantis} 
P. Antonioli et al., Nuclear Instrm. Methods  {\bf A433} 104-120, (1999). 


\bibitem{freund} 
M. Freund, Phys.Rev. D64 (2001) 053003; 
M. Freund, P. Huber, M. Lindner, Nucl.Phys. B615 (2001) 331-357;
 

\bibitem{barger} 
 V.D. Barger, S. Geer, R. Raja, K. Whisnant,  Phys. Rev. D63: 113011 (2001); 
V. Barger et al., 
hep-ph/0103052; P. Huber, M. Lindner, W. Winter, Nucl. Phys. B645, 3 (2002);
V. Barger, D. Marfatia, K. Whisnant,  Phys. Rev. D65: 073023 (2002). 























  









\end{thebibliography}
\end{document}